\documentclass[letterpaper,twocolumn,10pt]{article}
\usepackage{usenix2024_SOUPS}

\usepackage{tikz}
\usepackage{amsmath}
\usepackage{amssymb}
\usepackage{graphicx}
\usepackage{subcaption}

\begin{document}

\date{}

\title{\Large \bf PersonaFingerprint: Measuring Persona Inference on Modern Websites with LLM-Driven Browsing}


\def\plainauthor{Author name(s) for PDF metadata. Don't forget to anonymize for submission!}


\author{
{\rm Chuxu Song}\\
Rutgers University
\and
{\rm Hao Wang}\\
Rutgers University
\and
{\rm Richard Martin}\\
Rutgers University
}

\maketitle

\begin{abstract}
Website Fingerprinting (WFP) has traditionally focused on inferring \emph{which} website a user visits from encrypted traffic metadata such as packet sizes and timing. In this paper, we identify and quantify a new privacy risk in modern-web settings: an adversary can infer a user’s \emph{persona} using only packet-length and inter-arrival-time sequences. To study this risk at scale, we build an LLM-driven multi-agent browsing framework that enforces controllable persona constraints while a computer-use agent interacts with real websites and collects corresponding encrypted traffic traces. We formalize \textit{persona fingerprinting} under both closed-set and open-world settings and further evaluate whether persona information is already embedded in representations learned by existing WFP models and can be amplified at low cost. Across 10 modern websites and 15 personas (plus an open-world class), persona inference achieves about 84\% accuracy on mixed-site traffic; 
moreover, a lightweight multi-task objective can boost persona accuracy to around 80\% while retaining strong site classification performance (about 93\% baseline). Our results show that, on modern websites, encrypted traffic metadata can leak not only which site a user visits, but also how they browse and who is browsing.
\end{abstract}

\section{Introduction}
\label{sec:introduction}

Web traffic today is dominated by \emph{modern websites} that are highly interactive and continuously updated, including e-commerce platforms, video streaming services, social and community forums, and professional networking portals. These sites interleave user-driven interactions with background content fetching (e.g., recommendations, feeds, and embedded resources), producing rich and evolving network traces during a browsing session. Even when transport-layer encryption and anonymization hide application payloads, network vantage points---for example, VPN providers, ISPs, or enterprise gateways---can still observe fine-grained \emph{packet-level metadata}.

A large body of work on \emph{website fingerprinting}\cite{netclr,tor_wang,10179464,ares,Bhat_2019,277132,Rahman_2020,10.1145/3243734.3243832,raptor2017early,deng2025singleperspectiverealisticevaluation,10.1145/3319535.3354217} (WFP) has demonstrated that such metadata often suffices to infer which site a user is visiting through an encrypted or anonymized channel. In this paper, we investigate a complementary and, arguably, more sensitive question: \emph{to what extent can packet-level metadata reveal who a user is---that is, their behavioral persona---even when different users visit the same set of sites?} Rather than only predicting site identities, we study \textbf{persona fingerprinting}: given a short window of encrypted traffic, can an attacker distinguish between different personas based solely on browsing style? Examples include price-sensitive comparison shoppers, habitual short-video consumers, cautious information seekers who read in depth before clicking, and fast skimmers who rapidly navigate recommended content. Importantly, we assume that the attacker has no access to payloads, identifiers, cookies, or browser fingerprints, and only observes packet metadata.

Modern websites blur traditional request and page boundaries: asynchronous updates, lazy loading, and multiplexed resources make it unrealistic to assume an attacker can perfectly segment traffic into individual page loads\cite{oliveira2024browsingbehaviorexposesidentities,255662,panchenko2016website}. 
However, locating page boundaries on modern websites is unrealistic: many sites now use single-page application designs, video streaming, and infinite scrolling. We therefore adopt a fixed-length packet-window model\cite{song2025redefiningwebsitefingerprintingattacks,song2024seamlesswebsitefingerprintingmultiple,10001054}.
The attacker observes encrypted traffic metadata from a browsing session to a website (including embedded third-party resources such as ads), and operates on windows of 1{,}000 consecutive packets represented only by packet lengths (with direction) and inter-arrival times. The attacker does not know how these packets map to specific requests or objects and cannot rely on application-layer markers. The central question is whether such fixed-length packet windows contain enough structure to support not only website fingerprinting, but also persona fingerprinting.

Studying persona fingerprinting in this setting raises two main challenges. First, collecting large-scale, labeled traffic from human users is difficult: it requires extensive instrumentation, careful consent and ethics procedures, and typically yields limited coverage of the long-tail diversity of browsing behaviors. Second, even with sufficient data, it is unclear whether models trained for site-level tasks (such as WFP) already encode persona-level information in their internal representations, or whether persona fingerprinting requires fundamentally different features and architectures. At the same time, recent LLM-based agents\cite{zhou2024webarenarealisticwebenvironment,deng2023mind2webgeneralistagentweb,dechezelles2025browsergymecosystemwebagent,li2023camelcommunicativeagentsmind} change the landscape: if persona-conditioned browsing traces can be generated at scale, the barrier for attackers to obtain labeled behavioral data may drop substantially, strengthening persona inference from encrypted metadata.

To address these challenges, we leverage recent advances in large language model (LLM) agents to construct a controllable, multi-persona browsing framework\cite{song2025redefiningwebsitefingerprintingattacks,Wang_2024,sager2025comprehensivesurveyagentscomputer} that operates on real websites while avoiding direct collection of sensitive user data. Our framework couples two agents. A \textbf{persona-conditioned decision agent}, implemented as an LLM, is prompted with a structured persona profile and, at each step, receives a screenshot of the current page plus a compact history of past actions. Conditioned on this context, it produces the next high-level instruction in natural language (e.g., ``scroll through product reviews'', ``search for alternative brands'', ``open the second recommended video''). A \textbf{computer-use agent} then executes these instructions in a real browser, translating them into concrete DOM interactions (clicks, keystrokes, scrolling) and driving the underlying network traffic. The browser is instrumented with packet capture, allowing us to record the resulting encrypted traffic at packet granularity.

We instantiate fifteen distinct personas plus an \emph{open-world} (OW) category and run them across ten popular services spanning e-commerce, video, news, community Q\&A, and professional networking (including Amazon, YouTube, Reddit, CNN, Yelp, Bilibili, eBay, Yahoo, Zhihu, and LinkedIn). We collect fixed-length windows of 1{,}000 consecutive packets as training and evaluation samples. A site-only classifier achieves around 93\% average website fingerprinting accuracy across these ten sites, confirming that short packet windows remain sufficient for modern WFP. More importantly, we find that personas are also highly learnable from the same metadata: in a mixed-site setting, persona inference reaches about 84\% accuracy (15 personas plus OW). We further show that persona signals can be \emph{incidentally} encoded by WFP representations: freezing a site-only encoder and attaching a lightweight MLP probe\cite{belinkov2021probingclassifierspromisesshortcomings,ruder2017overviewmultitasklearningdeep} recovers persona accuracy far above random-feature baselines (roughly a 20--30 point absolute gain across several sites). These results highlight that encrypted traffic metadata can leak not only \emph{which} site a user visits, but also \emph{how} they browse.

Our goal is not to claim that LLM agents perfectly mimic all aspects of human behavior. Instead, we treat them as a controlled and reproducible mechanism to explore whether realistic, persona-conditioned browsing patterns can be distinguished at the level of packet metadata, and how scalable persona generation could impact privacy risk. We design persona profiles with explicit structure (background, goals, content preferences, interaction style) and encourage both \emph{persona consistency} (coherent, recognizable behavior under the same persona) and \emph{behavioral diversity} (multiple navigation paths and content choices) so that samples do not collapse into a single scripted trace.

Using this framework, we make the following contributions:
\begin{itemize}
  \item We introduce and formalize \textbf{persona fingerprinting} as a new privacy risk on \textbf{modern websites}, where an attacker---under the standard WFP vantage point (e.g., Tor/VPN adversaries)---attempts to infer behavioral personas from encrypted traffic metadata.
  \item We present a \textbf{multi-persona LLM browsing framework} that couples a persona-conditioned decision agent with a computer-use agent to generate packet-level traces on real websites at scale, supporting fifteen personas plus an open-world category without recruiting large numbers of human participants.
  \item We empirically evaluate persona fingerprinting under fixed-length \textbf{packet windows}. Given only 1{,}000-packet windows (packet lengths with direction and inter-arrival times), our classifiers achieve strong persona classification performance across multiple services; on five representative platforms---Bilibili, eBay, Yahoo, Zhihu, and LinkedIn---per-site macro-F$_1$ typically ranges from approximately 0.78 to 0.91, and mixed-site persona inference achieves about 84\% accuracy.
  \item We quantify \textbf{incidental persona leakage} in WFP models via representation probing: a frozen site-only encoder with a lightweight MLP probe substantially improves persona inference over random-feature baselines (about 20--30 points absolute), illustrating how ostensibly site-focused traffic models may be repurposed into persona fingerprinting tools with minimal additional effort.
\end{itemize}

Overall, our study highlights that who a user is online may be learnable from encrypted traffic even when the attacker observes only fixed-length packet windows and no application payloads. Persona fingerprinting thus deserves attention alongside traditional website fingerprinting in the design and evaluation of privacy-preserving web systems.

\section{Threat Model and Problem Definition}
\label{sec:threat}

We adopt the standard threat model used in website fingerprinting (WFP): a passive network adversary\cite{panchenko2016website,281438,Abe2016FingerprintingAO} who observes encrypted traffic on anonymity or tunneling channels such as Tor or VPNs. This vantage point is representative of entities such as VPN providers, ISPs, or enterprise gateways, which can monitor packet-level metadata but cannot decrypt application payloads. Throughout, we assume the adversary has access only to \emph{traffic metadata}---packet lengths (with direction) and inter-arrival times---and has no access to URLs, page contents, cookies, browser identifiers, or other client-side signals.

Crucially, we focus on \emph{modern websites} where interactive browsing is interleaved with continuous background fetching (e.g., feeds, recommendations, and embedded resources). Accordingly, the adversary observes encrypted traffic metadata from a \emph{browsing session} to a target website, including naturally embedded third-party resources such as ads. The adversary does not know how packets map to individual requests or objects, and cannot rely on application-layer markers to segment page loads.

\subsection{Observation and Representation}
\label{sec:representation}
Because asynchronous updates and continuous loading on modern websites blur traditional page boundaries, we model the adversary's observation using fixed-length \emph{packet windows}, a common abstraction in modern traffic analysis. Each sample is represented as a sequence
\begin{equation}
X = \{(l_t, d_t)\}_{t=1}^{L},
\end{equation}
where $l_t$ denotes the signed packet length (with the sign encoding direction) and $d_t$ denotes the inter-arrival time between packet $t-1$ and $t$. We set the window length to $L=1000$ packets.

We extract \emph{non-overlapping} packet windows \emph{within a single connection} and discard trailing segments shorter than $L$ packets. Unless otherwise stated, all learning and evaluation in this paper operates on these fixed-length packet windows and uses only $(l_t, d_t)$ metadata.

\subsection{Problem Definition}
Let $\mathcal{S}$ denote the set of monitored websites, with $|\mathcal{S}|=10$ in our study. Let $\mathcal{P}$ denote the set of canonical behavioral personas, with $|\mathcal{P}|=15$. In addition, we introduce an \emph{open-world} label, denoted $\mathsf{OW}$, which aggregates multiple \emph{unseen} personas (i.e., browsing styles not covered by the canonical set). We use $\mathsf{OW}$ to capture \emph{misattribution risk}: when a user does not match any canonical persona, the adversary may still incorrectly assign them to a specific known persona. In our datasets, the open-world category is constructed per site and balanced to be comparable in scale to canonical personas.

\paragraph{Website fingerprinting (WFP).}
Given a packet window $X$, the WFP task is to predict the visited website:
\begin{equation}
f_{\text{site}}: X \rightarrow s,\quad s \in \mathcal{S}.
\end{equation}

\paragraph{Persona fingerprinting.}
Given a packet window $X$, the persona fingerprinting task is to infer the user's behavioral persona:
\begin{equation}
f_{\text{pers}}: X \rightarrow p,\quad p \in \mathcal{P} \cup \{\mathsf{OW}\}.
\end{equation}
We consider both \emph{closed-set} classification over $\mathcal{P}$ and \emph{open-world} classification over $\mathcal{P}\cup\{\mathsf{OW}\}$.

\paragraph{Per-site vs.\ mixed-site settings.}
We evaluate persona fingerprinting under two complementary settings that reflect different attacker assumptions. In the \emph{per-site} setting, we condition on a target website $s$ (e.g., by prior knowledge or a separate WFP stage) and ask whether the adversary can distinguish personas \emph{within} that site. In the \emph{mixed-site} setting, we do not rely on site identity and instead infer personas from traffic pooled across websites, assessing whether persona signals are robust and site-agnostic under heterogeneous modern-web traffic.

\subsection{Metrics}
\label{sec:metrics}
We quantify adversarial success using standard classification metrics, primarily accuracy and macro-F$_1$, which serve as empirical estimates of the risks defined above. In open-world experiments, we additionally examine the behavior of the $\mathsf{OW}$ label to characterize misattribution, i.e., whether unseen personas are incorrectly mapped to one of the canonical personas.

\section{LLM-Driven Persona Browsing for Controlled Traffic Generation}
\label{sec:methodology}

This section describes how we generate persona-conditioned browsing behavior on \emph{modern websites} using an LLM-driven multi-agent framework, and how we obtain labeled packet windows suitable for both website fingerprinting and \emph{persona fingerprinting}. Our focus is the generation process and its quality controls. Beyond enabling controlled measurement, scalability is pivotal from a security perspective: LLM agents can produce large amounts of persona-labeled browsing traces, which can substantially lower the barrier for adversaries to train powerful metadata-based inference models. Section~\ref{sec:modeling} then describes the downstream models used to attack the resulting packet windows.

\subsection{Framework Overview}
Our goal is to simulate realistic, persona-conditioned browsing on real websites while retaining full control over labels and avoiding direct collection of sensitive user content. We deploy a two-agent framework on top of a real browser and instrument the network interface to capture encrypted traffic metadata.

For each website $s \in \mathcal{S}$ and persona label $p \in \mathcal{Y}_{\mathsf{pers}}$ (fifteen canonical personas plus an open-world label), we (i) launch a fresh browser instance and navigate to the entry page of $s$; (ii) run a multi-step interaction loop between a \emph{persona-conditioned decision agent} (an LLM) and a \emph{computer-use agent} that controls the browser; (iii) continuously capture packets observed during the session; and (iv) segment captured traces offline into non-overlapping packet windows of length $L{=}1000$, each inheriting the website label $s$ and persona label $p$. As in our threat model (Section~\ref{sec:threat}), a browsing session includes the website's naturally embedded third-party resources (e.g., ads). Importantly, we do \emph{not} concatenate traffic across connections: packet windows are extracted \emph{within a single connection}, and we discard trailing segments shorter than $L$ packets.

We repeat this procedure to build a large labeled corpus; in our experiments, each (site, persona) pair yields approximately 5{,}000 packet windows, resulting in roughly $10 \times 16 \times 5{,}000 \approx 800{,}000$ labeled windows. Segmentation into fixed-length windows is a purely \emph{offline} step that standardizes the learning and evaluation unit under our packet-window setting.

\subsection{Persona Design and Open-World Construction}
\label{sec:persona-design}

\paragraph{Canonical personas.}
We design fifteen canonical personas $\mathcal{P} = \{p_1,\dots,p_{15}\}$ intended to capture diverse yet plausible browsing styles. Each persona is specified via a structured natural-language profile with three components:
\begin{itemize}
  \item \textbf{Persona profile.} Background attributes (e.g., profession, technical experience), typical goals (e.g., bargain hunting, quick entertainment, in-depth research, job seeking), and risk attitudes (e.g., cautious about unfamiliar links).
  \item \textbf{Session pattern.} Preferred interaction style and pace, such as reliance on search versus recommendations, how deeply they scroll, how often they open new tabs, and how long they linger on pages.
  \item \textbf{Content preferences and avoidance.} High-level interests and aversions, such as prioritizing user reviews over marketing copy, preferring long-form articles versus snippets, or avoiding clickbait.
\end{itemize}
These components form a persona-specific system prompt provided to the decision agent. Each canonical persona is instantiated on \emph{every} website in $\mathcal{S}$, enabling both per-site and mixed-site persona fingerprinting evaluation.

\paragraph{Open-world label.}
In addition to the canonical personas, we define a single open-world label $\mathsf{OW}$ that aggregates multiple \emph{unseen} personas (i.e., browsing styles not covered by $\mathcal{P}$). Concretely, we construct a set of additional persona profiles and generate a small amount of browsing traffic for each; all such traffic is folded into the $\mathsf{OW}$ label. We balance $\mathsf{OW}$ \emph{per site} to be comparable in scale to canonical personas, allowing us to quantify \emph{misattribution risk}: when a user does not match any canonical persona, does an adversary incorrectly assign them to a specific known persona?

\subsection{Decision Loop: Decision Agent and Computer-Use Agent}
\label{sec:decision-loop}

Our framework couples a persona-conditioned LLM decision agent with a computer-use agent that executes actions in a real browser. Time is divided into abstract decision steps $t = 1,2,\dots$.

\paragraph{Inputs to the decision agent.}
At each step $t$, the decision agent receives (i) the persona system prompt for $p$ (fixed throughout the session); (ii) a representation of the current page, primarily a screenshot; (iii) a compact natural-language history summary of recent actions and outcomes; and (iv) global constraints (remain on the current domain; avoid login/payment flows; do not input sensitive information).

Conditioned on this input, the decision agent generates a high-level natural-language instruction $a_t$ (e.g., ``scroll and read several critical reviews'' or ``use search to compare alternatives and open a detailed result'').

\paragraph{Action normalization and constraints.}
The instruction $a_t$ passes through a lightweight normalization layer that maps it to a structured action from a finite action set (e.g., \textsc{Search}$(\textit{text})$, \textsc{Click}$(\textit{selector})$, \textsc{Scroll}$(\textit{direction, amount})$, \textsc{Navigate}$(\textit{section})$). We enforce safety and experimental constraints (no login, passwords, or payments; no navigation outside the target domain). If a violation is detected, we rewrite the instruction into a compliant alternative.

\paragraph{Execution and logging.}
The computer-use agent executes the structured action via DOM interactions (clicks, typing, scrolling, tab management) and waits for the page to stabilize. It then captures the next screenshot, updates the history summary, and logs coarse behavioral features (action type, dwell time, page type, tab events) used for later sanity checks. Packet capture runs continuously during the session; segmentation into packet windows is performed offline as described above.

\subsection{Quality Controls: Persona Consistency and Behavioral Diversity}
\label{sec:quality}

Persona-conditioned generation must satisfy two competing goals: (i) \emph{consistency}, meaning behavior should align with the persona description; and (ii) \emph{diversity}, meaning sessions should not collapse into a rigid script. We quantify both properties using lightweight, interpretable metrics that support the plausibility of generated traffic independent of downstream classifier accuracy.

\subsubsection{Persona Consistency}
\label{sec:consistency}

\paragraph{Trajectory summaries.}
For a trajectory segment $\tau$ consisting of several decision steps (e.g., 5--10 high-level actions), we construct a short natural-language summary $T_\tau$ from the logged action sequence and page types (via simple templates or a small auxiliary LLM). Let $T_p$ denote the persona description text for persona $p$ (profile + session pattern + content preferences).

\paragraph{Text-based alignment score.}
We define a persona consistency score $C(p,\tau)\in[0,1]$ measuring how well $T_\tau$ matches $T_p$. In practice, we instantiate $C(p,\tau)$ using either (i) normalized cosine similarity between sentence embeddings $E(T_p)$ and $E(T_\tau)$, or (ii) an LLM-based rating prompt mapped to $[0,1]$. For each persona $p$, we report an average consistency score
\[
C(p) = \mathbb{E}_{\tau \sim \text{traj}(p)}\!\left[ C(p,\tau) \right].
\]
We use $C(p,\tau)$ as a sanity check that persona prompts influence behavior as intended, and optionally to discard or downweight segments that are clearly out-of-character.

\subsubsection{Behavioral Diversity}
\label{sec:diversity}

Consistency alone does not prevent degenerate scripted behavior. We therefore measure diversity at both the action level (what the agent does) and the packet level (what an attacker observes).

\paragraph{Action-level diversity.}
Let $p_p(a)$ be the empirical frequency of action type $a$ across all trajectories generated under persona $p$. We define the action entropy
\[
H_{\mathrm{act}}(p) = - \sum_a p_p(a)\log p_p(a),
\]
where higher entropy indicates a richer mix of actions. We also measure coarse site coverage by abstracting page states into a finite set of types (e.g., homepage, search results, detail page, comments/reviews, feed/recommendations). Let $\text{VisitedStates}(p,s)$ be the set of page types visited by persona $p$ on site $s$, and let $\text{AllStates}(s)$ be the union across all personas. We define
\[
\mathrm{Coverage}(p,s) = \frac{|\text{VisitedStates}(p,s)|}{|\text{AllStates}(s)|}.
\]
We combine these into an action-level diversity score
\[
D_{\mathrm{act}}(p) = \alpha H_{\mathrm{act}}(p) + (1-\alpha)\,\mathbb{E}_{s}\!\left[\mathrm{Coverage}(p,s)\right],
\]
with $\alpha\in[0,1]$ controlling the tradeoff between action richness and functional coverage.

\paragraph{Packet-level diversity.}
Action diversity may not fully reflect diversity in the packet sequences observed by an adversary. Let $h(\cdot)$ be an encoder mapping packet windows $X$ to a latent representation in $\mathbb{R}^d$ (instantiated by the encoders described in Section~\ref{sec:modeling}). For persona $p$, let $\mathcal{X}_p$ be the set of all packet windows labeled with $p$. We define the persona mean
\[
\mu_p = \frac{1}{|\mathcal{X}_p|}\sum_{X\in\mathcal{X}_p} h(X),
\]
and the intra-persona spread
\[
D_{\mathrm{pkt}}(p) = \frac{1}{|\mathcal{X}_p|}\sum_{X\in\mathcal{X}_p} \left\|h(X)-\mu_p\right\|_2^2.
\]
Intuitively, $D_{\mathrm{pkt}}(p)$ is low for near-identical packet patterns (indicative of a rigid script) and increases as packet-level behavior explores a broader region of the representation space. In our evaluation, we report both action-level and packet-level diversity to show that personas are coherent yet non-degenerate.



\section{Modeling Website and Persona Fingerprints}
\label{sec:modeling}

This section describes how an adversary can learn both website and persona fingerprints from the packet-window representation defined in Section~\ref{sec:representation}, under the threat model in Section~\ref{sec:threat}. We consider a shared packet-window encoder that maps encrypted traffic metadata to a compact representation, and attach lightweight prediction heads for (i) website fingerprinting (WFP) and (ii) persona fingerprinting. We further study two complementary mechanisms by which persona information may arise in practice: \emph{(a)} direct training for persona inference, and \emph{(b)} \emph{incidental leakage} where WFP-focused representations become repurposable for persona inference with minimal additional effort. For notational convenience, we denote the full persona label set by
\[
\mathcal{Y}_{\mathsf{pers}} \;=\; \mathcal{P} \cup \{\mathsf{OW}\},
\]
where $\mathcal{P}$ is the canonical persona set and $\mathsf{OW}$ is the open-world label (Section~\ref{sec:threat}).

\subsection{Shared Packet-Window Encoder}
\label{sec:encoder}

Let $X$ denote a fixed-length packet window (Section~\ref{sec:representation}) consisting only of packet lengths (with direction) and inter-arrival times. We model $X$ using a neural encoder
\[
h_\theta: X \rightarrow \mathbf{z}\in\mathbb{R}^d,
\]
which produces a $d$-dimensional representation $\mathbf{z}$ for each window. In our implementation, $h_\theta(\cdot)$ is instantiated as a 1D CNN sequence encoder over the per-packet metadata sequence, followed by a pooling operator to obtain a single vector for the window. We intentionally keep the encoder generic: our goal is not to propose a new architecture, but to examine what \emph{information} about websites and personas can be learned from encrypted traffic metadata under modern-web browsing.

\subsection{Website Fingerprinting Head}
\label{sec:wfp-head}

We model standard WFP with a softmax classifier on top of the shared encoder. Given $\mathbf{z}=h_\theta(X)$, the website predictor is
\[
f_{\mathsf{site}}(X) \;=\; \mathrm{softmax}(W_s \mathbf{z} + \mathbf{b}_s),
\]
where $W_s$ and $\mathbf{b}_s$ are learned parameters. Training minimizes the cross-entropy loss
\[
\mathcal{L}_{\mathsf{site}} \;=\; \mathbb{E}_{(X,s)\sim \mathcal{D}} \left[-\log f_{\mathsf{site}}(X)[s]\right],
\]
with $s\in \mathcal{S}$ as defined in Section~\ref{sec:threat}. This site-only model serves as both a baseline for modern WFP performance and a foundation for analyzing whether representations learned for WFP can \emph{incidentally} encode persona information.

\subsection{Persona Fingerprinting Heads}
\label{sec:persona-heads}

Persona fingerprinting aims to infer a behavioral persona label $p\in \mathcal{Y}_{\mathsf{pers}}$ from the same packet-window metadata. We attach a persona classifier to the shared encoder:
\[
f_{\mathsf{pers}}(X) \;=\; \mathrm{softmax}(W_p \mathbf{z} + \mathbf{b}_p), \qquad \mathbf{z}=h_\theta(X),
\]
and train using cross-entropy over persona labels. We consider two settings (Section~\ref{sec:threat}), which correspond to different attacker assumptions.

\paragraph{Per-site persona fingerprinting.}
In the per-site setting, the adversary conditions on a target website $s$ (e.g., by prior knowledge or a separate WFP stage) and seeks to distinguish personas \emph{within that site}. Formally, we learn a persona predictor under the conditional distribution $\mathcal{D}(X,p \mid s)$:
\[
\mathcal{L}_{\mathsf{pers}}^{\mathsf{(per\text{-}site)}} \;=\; 
\mathbb{E}_{(X,p)\sim \mathcal{D}(\cdot\mid s)} \left[-\log f_{\mathsf{pers}}(X)[p]\right].
\]
This captures the strongest attacker setting: the website is fixed, so persona inference must rely on behavior-induced differences in packet metadata rather than site-level structure.

\paragraph{Mixed-site persona fingerprinting.}
In the mixed-site setting, the adversary does not rely on website identity and instead infers personas from packet windows pooled across websites:
\[
\mathcal{L}_{\mathsf{pers}}^{\mathsf{(mixed)}} \;=\; 
\mathbb{E}_{(X,p)\sim \mathcal{D}} \left[-\log f_{\mathsf{pers}}(X)[p]\right].
\]
This setting is more stringent: it evaluates whether persona signals are robust and \emph{site-agnostic} under heterogeneous modern-web traffic. Intuitively, success here suggests that persona-related behavioral patterns (e.g., interaction cadence, dwell-and-scroll patterns, and navigation breadth) induce consistent packet-metadata signatures that generalize across sites.

\paragraph{Closed-set vs.\ open-world.}
We evaluate persona inference under both closed-set classification over $\mathcal{P}$ and open-world classification over $\mathcal{Y}_{\mathsf{pers}}=\mathcal{P}\cup\{\mathsf{OW}\}$ (Section~\ref{sec:threat}). The open-world label $\mathsf{OW}$ enables us to quantify \emph{misattribution risk}: whether traffic from unseen personas is incorrectly assigned to a specific canonical persona, which can amplify privacy harms by turning ``unknown'' behavior into a concrete (and potentially sensitive) inferred profile.

\subsection{Joint Multi-Task Learning}
\label{sec:joint}

In practice, an attacker may simultaneously pursue both WFP and persona inference, or may start from a WFP model and extend it with persona labels once such labels become available at scale (e.g., via automated persona-conditioned browsing as in Section~\ref{sec:methodology}). To model this low-friction upgrade path, we consider joint multi-task learning with a shared encoder and two heads:
\[
\mathcal{L}_{\mathsf{joint}} \;=\; \mathcal{L}_{\mathsf{site}} \;+\; \lambda \,\mathcal{L}_{\mathsf{pers}},
\]
where $\lambda \ge 0$ balances site and persona objectives. Joint training makes two phenomena explicit. First, it captures \emph{amplification}: even a lightweight additional objective can encourage the shared representation to preserve persona-discriminative cues that might otherwise be underutilized. Second, it quantifies the potential \emph{trade-off}: as $\lambda$ increases, the representation may allocate more capacity to persona variation at the expense of site classification. In Section~\ref{sec:evaluation}, we sweep $\lambda$ to empirically characterize this attack-relevant trade-off between ``which site'' and ``how the user browses.''

\subsection{Probing Persona Leakage in WFP Representations}
\label{sec:probing}

A central question is whether persona information is already present in representations learned for WFP, even when the model is trained \emph{only} for website classification. To measure such \emph{incidental leakage}, we use a two-stage probing protocol.
A \emph{probe} is a lightweight classifier trained on top of \emph{frozen} representations to test whether a target attribute (persona) is readily decodable from the learned features.

\paragraph{Stage 1: Train a site-only encoder.}
We train the encoder $h_\theta$ and site head $f_{\mathsf{site}}$ to minimize $\mathcal{L}_{\mathsf{site}}$ (Section~\ref{sec:wfp-head}) and obtain the trained encoder $h_{\theta^\star}$. We then \emph{freeze} $h_{\theta^\star}$.

\paragraph{Stage 2: Train a lightweight MLP probe for personas.}
We attach a shallow MLP probe on top of the frozen representation $\mathbf{z}=h_{\theta^\star}(X)$:
\[
\mathbf{u} = \sigma(W_1 \mathbf{z} + \mathbf{b}_1), \qquad
f_{\mathsf{probe}}(X) = \mathrm{softmax}(W_2 \mathbf{u} + \mathbf{b}_2),
\]
and train only the probe parameters $\{W_1,\mathbf{b}_1,W_2,\mathbf{b}_2\}$ using cross-entropy on persona labels. Importantly, the probe is deliberately low-capacity relative to the encoder; thus, strong probe performance indicates that persona cues are \emph{readily decodable} from WFP representations without extensive additional modeling.

\paragraph{Controls for fairness.}
To ensure that improvements are attributable to information encoded by $h_{\theta^\star}$ (rather than probe capacity), we train the \emph{same} MLP probe (with identical architecture and training budget) on top of a \emph{randomly initialized} frozen encoder as a control. The performance gap between the WFP-trained encoder and the random encoder provides an empirical lower bound on persona information embedded in WFP representations. This protocol reflects a realistic risk: once a strong WFP encoder exists, an attacker may repurpose it for persona inference with minimal additional effort and limited labeled persona data.

In summary, the models in this section operationalize three attacker-relevant pathways: (i) direct persona classifiers, (ii) multi-task amplification that strengthens persona inference as an incremental extension of WFP, and (iii) probing that quantifies incidental persona leakage already present in WFP representations. In Section~\ref{sec:evaluation}, we evaluate these pathways under both per-site and mixed-site settings and under closed-set and open-world persona definitions.

\section{Evaluation}
\label{sec:evaluation}

Section~\ref{sec:modeling} described three attacker-relevant pathways: (i) direct persona classifiers trained on packet windows, (ii) joint multi-task learning that upgrades a WFP encoder to also optimize persona inference, and (iii) probing that quantifies incidental persona leakage already present in WFP representations. We now evaluate these pathways under both per-site and mixed-site settings, and under closed-set and open-world persona definitions. In addition, we report behavioral diagnostics for LLM-generated traces and quantify how attack accuracy scales with the amount of persona-labeled traffic.

\subsection{Experimental Setup}


\paragraph{Websites and personas.}
We target ten popular websites spanning e-commerce, news, media, Q\&A, and professional networking:
\emph{Amazon}, \emph{CNN}, \emph{Yelp}, \emph{YouTube}, \emph{Reddit},
\emph{Bilibili}, \emph{eBay}, \emph{Yahoo}, \emph{Zhihu}, and \emph{LinkedIn}.
For each site we consider \(15\) canonical personas (Table~\ref{tab:personas}) plus an open-world label \(\mathsf{OW}\).
As defined in Section~\ref{sec:threat}, \(\mathsf{OW}\) aggregates multiple additional \emph{unseen} personas into a single ``other'' class (constructed per site and balanced to be comparable in scale to canonical personas), enabling us to quantify open-world behavior and misattribution risk.

\begin{table}[t]
\centering
\small
\setlength{\tabcolsep}{6pt}
\begin{tabular}{cl}
\toprule
\textbf{ID} & \textbf{Canonical persona} \\
\midrule
P1  & Time-Pressed Tech Professional \\
P2  & Deep-Dive Academic Researcher \\
P3  & Entertainment-Driven College Student \\
P4  & Cautious Older Non-Expert User \\
P5  & Price-Sensitive Online Shopper \\
P6  & Privacy-Conscious Security Enthusiast \\
P7  & Mobile-First Social Scroller \\
P8  & Task-Oriented Parent Planner \\
P9  & Comparison-Heavy Analyst \\
P10 & Loyal Brand/Topic Follower \\
P11 & Skeptical Fact-Checker \\
P12 & Map-and-Photo Local Explorer \\
P13 & Low-Patience Speed Skimmer \\
P14 & Accessibility-Careful Navigator \\
P15 & Curious Generalist Explorer \\
\bottomrule
\end{tabular}
\vspace{2pt}
\caption{\textbf{Canonical personas} used in closed-set experiments (P1--P15). We denote the open-world class by \(\mathsf{OW}\), which aggregates additional unseen personas (Section~\ref{sec:threat}).}
\label{tab:personas}
\end{table}

\noindent\textbf{Examples (intuition).}
P1 is goal-driven and time-pressed (fast search/navigation and short dwell),
P2 emphasizes deep reading and cross-checking (long dwell and multi-hop exploration),
P5 is cost-driven (frequent filtering/comparison and deal-seeking),
P4 is cautious and slow-paced (avoids risky clicks/popups and prefers stable paths),
and P13 exhibits very low patience (rapid skimming and frequent exits).
We refer to personas by their IDs (P1--P15) throughout figures and tables for readability.


\paragraph{Train/validation/test splits.}
To avoid temporal leakage, we split data at the level of long browsing sessions rather than individual windows: all windows from a given session go entirely to train, validation, or test. We use a \(70/10/20\) split and apply the \emph{same} split for all downstream tasks (website classification, persona classification, probing, and joint multi-task training), making results directly comparable.

\paragraph{Metrics.}
For website and persona classification we report top-1 accuracy and macro-\(F_1\) over classes. In the open-world persona setting, classes are the \(15\) canonical personas plus \(\mathsf{OW}\). To characterize open-world behavior beyond macro-\(F_1\), we additionally report \(\mathsf{OW}\) precision/recall/\(F_1\) and two attacker-relevant quantities: \textbf{MisAttr@OW} (the fraction of \(\mathsf{OW}\) windows predicted as any canonical persona) and \textbf{Top-3 Share} (how concentrated misattributions are among the top-3 predicted canonical personas). For behavioral diagnostics (Section~\ref{sec:evaluation-consistency}) we report the persona consistency score \(C(p)\) and site coverage defined in Section~\ref{sec:quality}. For joint multi-task training (Section~\ref{sec:evaluation-joint}) we study how website and persona accuracy evolve as a function of the persona loss weight \(\lambda\). Finally, to quantify the \emph{scalability} of attacks enabled by persona-labeled traffic, we evaluate how mixed-site open-world accuracy scales with the number of training windows per persona (Section~\ref{sec:evaluation-scaling}).

\subsection{Website Fingerprinting Baseline}
\label{sec:evaluation-site-only}

We first evaluate a website-only WFP baseline, which uses the shared encoder \(h_{\theta}(X)\) from Section~\ref{sec:modeling} with a website classification head \(f_{\mathsf{site}}\) and is trained \emph{only} on website labels, ignoring personas. This captures a realistic attacker who already deploys WFP and does not explicitly target personas.

\begin{figure}[t]
  \centering
  \includegraphics[width=\linewidth]{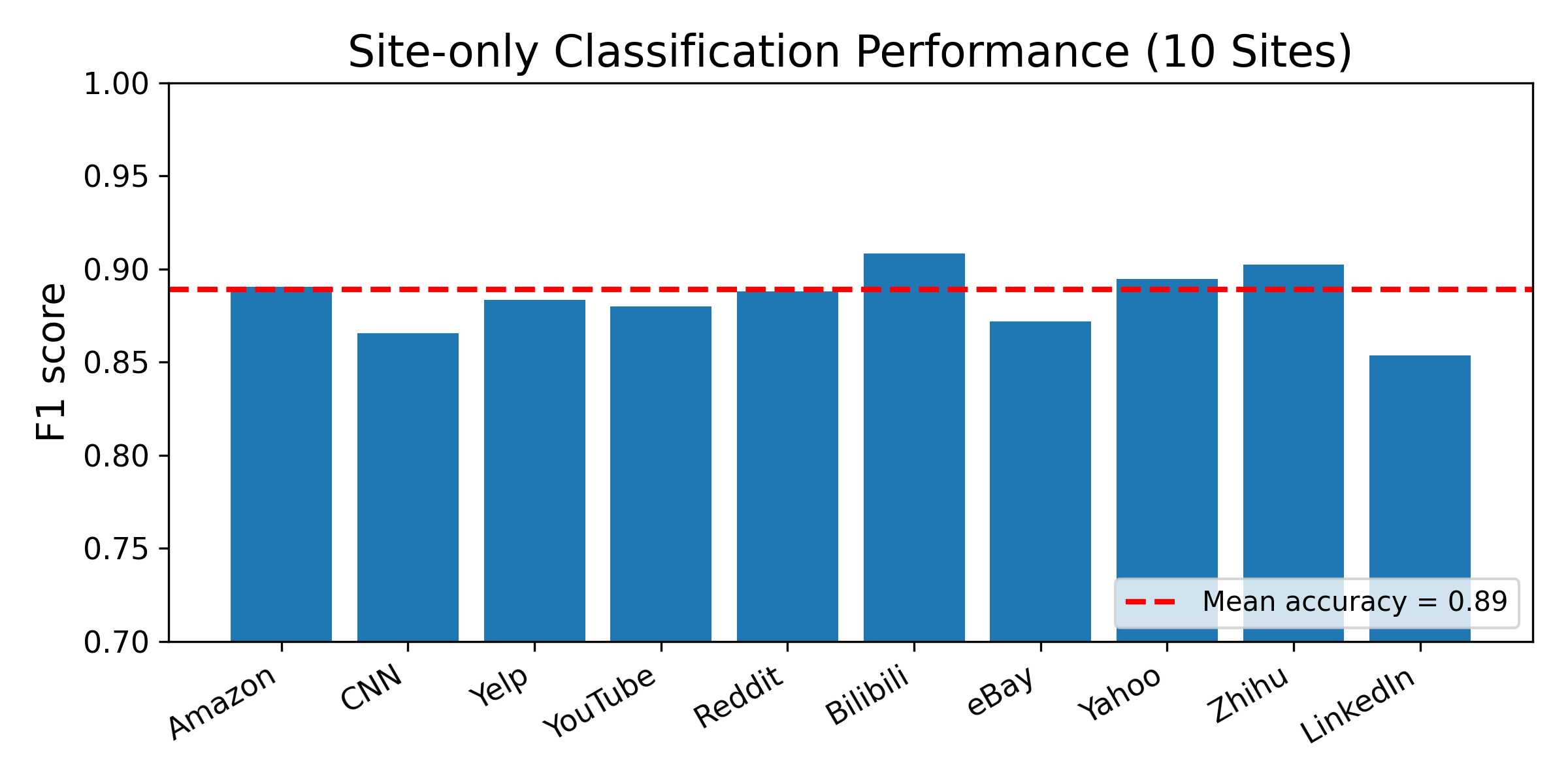}
  \caption{Website fingerprinting baseline: per-site accuracy and macro-\(F_1\) of the site-only classifier across all ten target websites using 1,000-packet windows. All sites achieve high accuracy (typically \(90\%\!-\!95\%\)), indicating that the encoder \(h_{\theta}\) learns strong discriminative features from packet metadata alone.}
  \label{fig:site-only-wfp}
\end{figure}

Figure~\ref{fig:site-only-wfp} shows per-site performance of this baseline. Across the original five English-language sites (Amazon, CNN, Yelp, YouTube, Reddit), top-1 accuracy averages around \(93\%\), with macro-\(F_1\) similarly above \(0.90\). Extending to all ten sites, accuracy for each website remains in the \(90\%\!-\!95\%\) range. This confirms that, under our 1{,}000-packet window representation, modern WFP models are already highly effective at identifying websites from encrypted traffic. In the next sections we show that (i) strong persona fingerprints are learnable and (ii) WFP encoders can be cheaply repurposed toward persona inference, making the upgrade path for an attacker particularly realistic.

\subsection{Per-Site Persona Fingerprinting}
\label{sec:evaluation-per-site-persona}

We now turn to persona classification \emph{within each site}. We train a persona classifier \(f_{\mathsf{pers}}^{(S)}\) for each target site \(S\), using the same encoder \(h_{\theta}\) but replacing the head with a \(16\)-way classifier for the fifteen canonical personas plus \(\mathsf{OW}\).

\begin{figure*}[t]
  \centering
  \begin{subfigure}{0.19\textwidth}
    \centering
    \includegraphics[width=\linewidth]{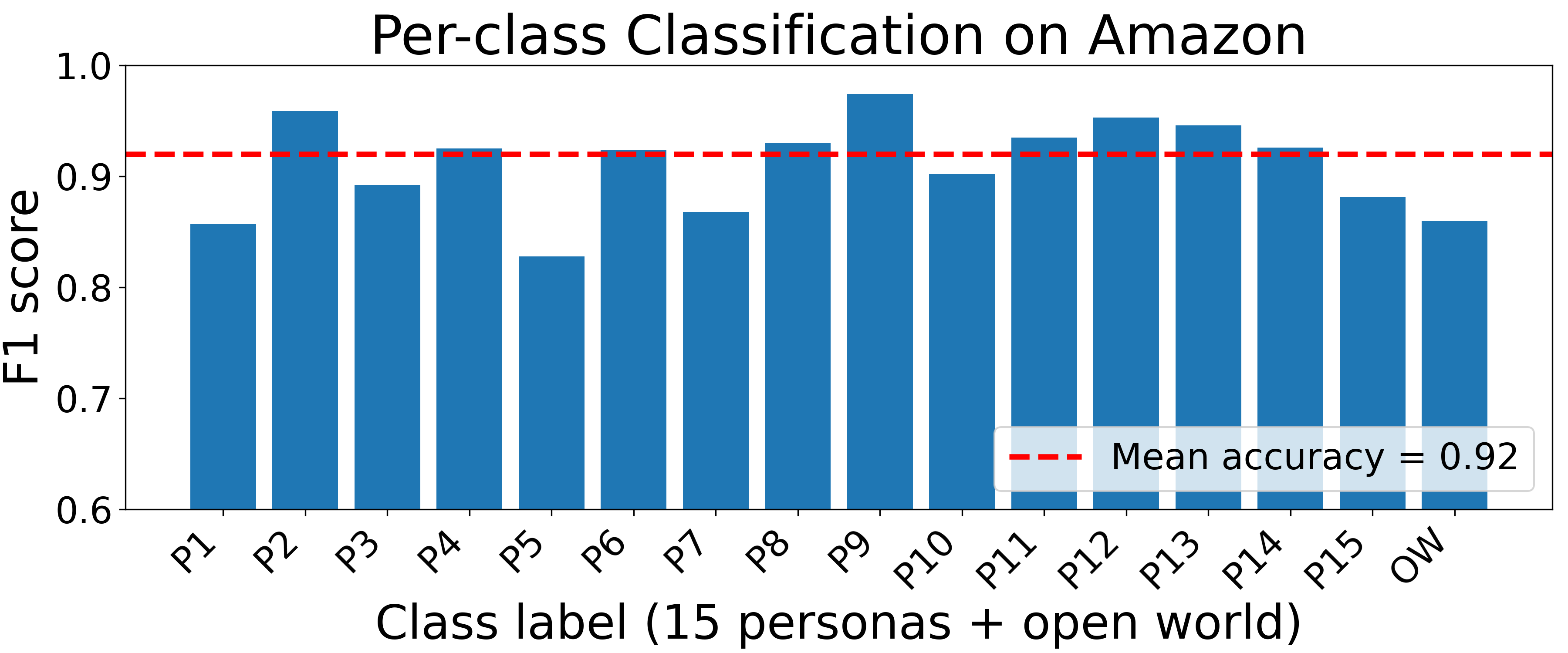}
    \caption{Amazon}
  \end{subfigure}
  \hfill
  \begin{subfigure}{0.19\textwidth}
    \centering
    \includegraphics[width=\linewidth]{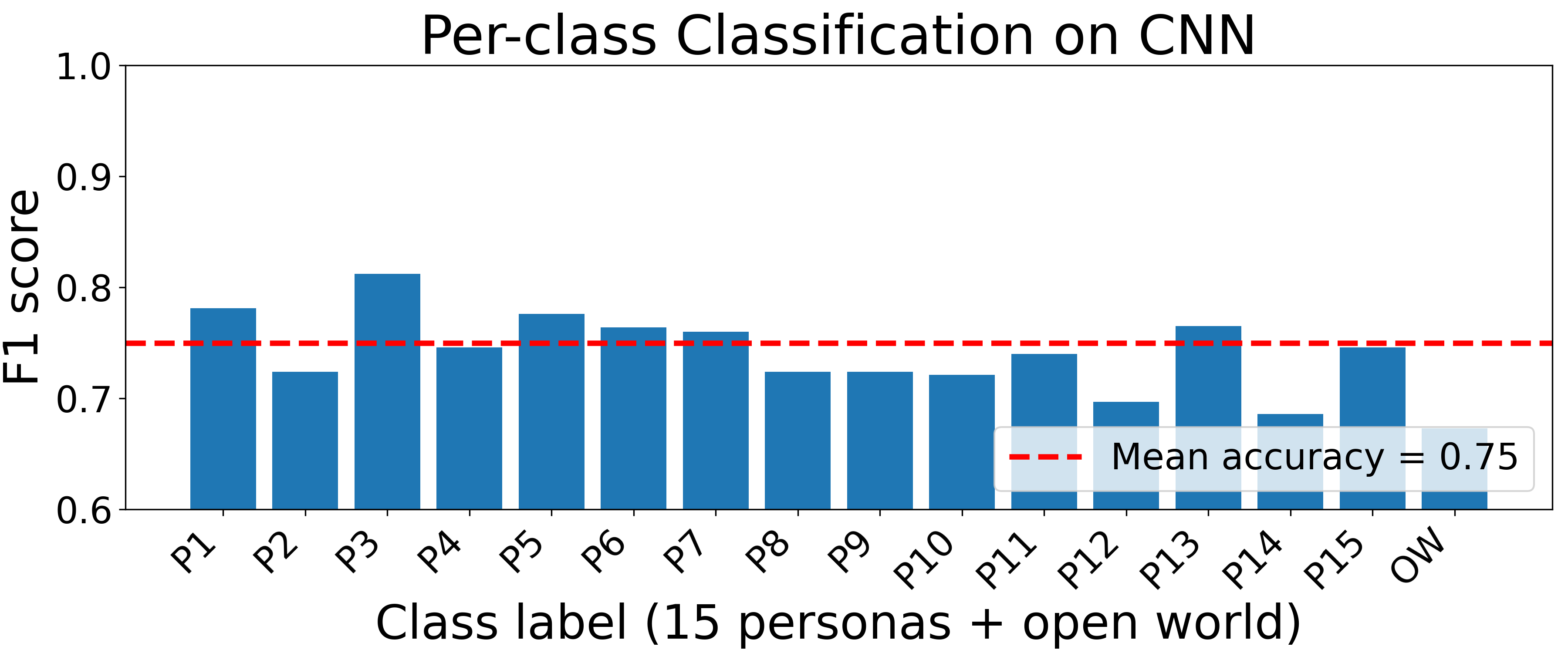}
    \caption{CNN}
  \end{subfigure}
  \hfill
  \begin{subfigure}{0.19\textwidth}
    \centering
    \includegraphics[width=\linewidth]{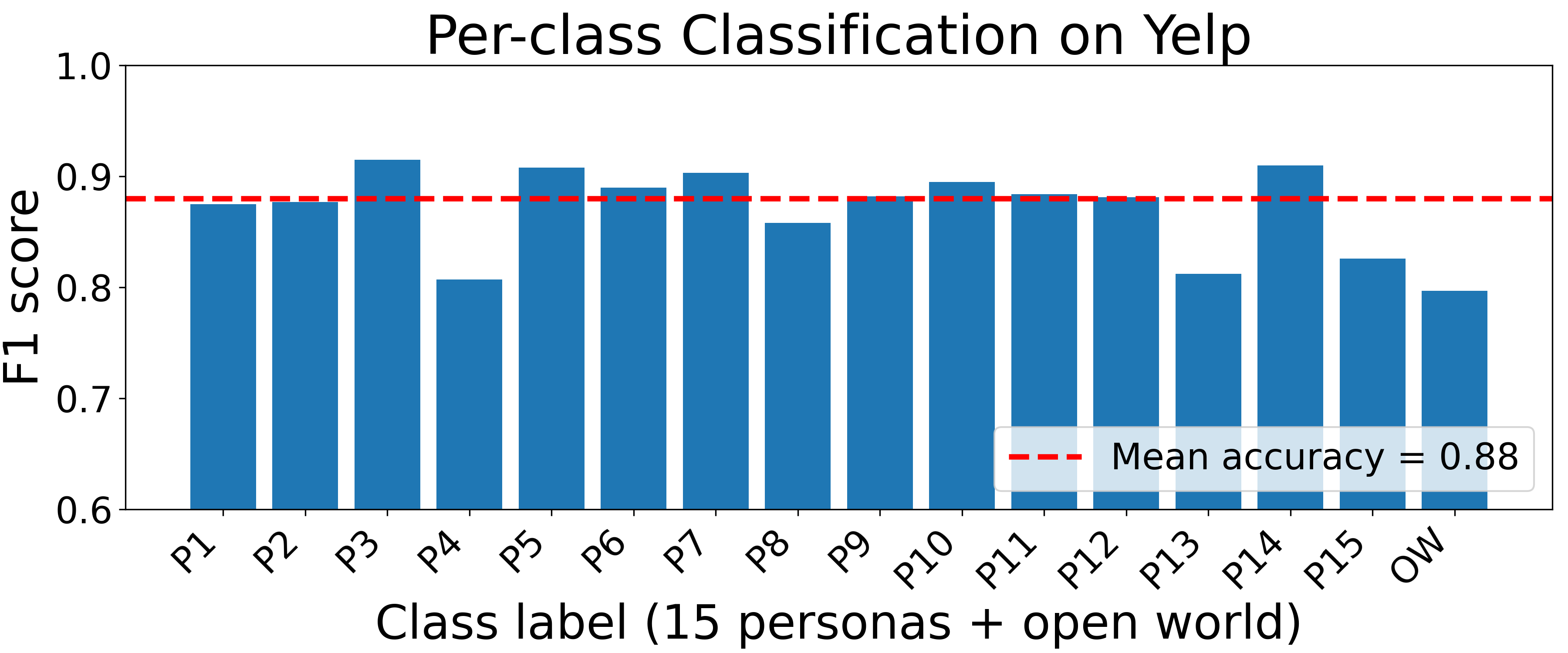}
    \caption{Yelp}
  \end{subfigure}
  \hfill
  \begin{subfigure}{0.19\textwidth}
    \centering
    \includegraphics[width=\linewidth]{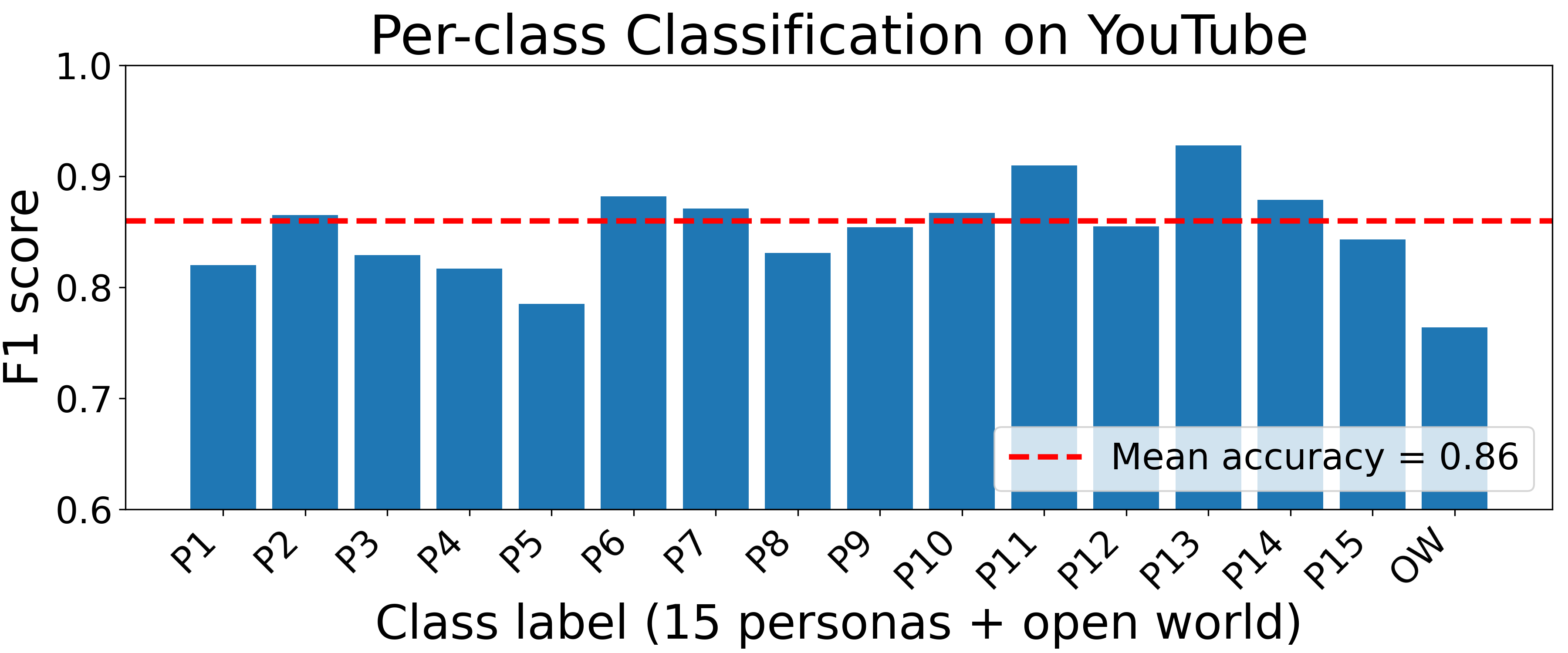}
    \caption{YouTube}
  \end{subfigure}
  \hfill
  \begin{subfigure}{0.19\textwidth}
    \centering
    \includegraphics[width=\linewidth]{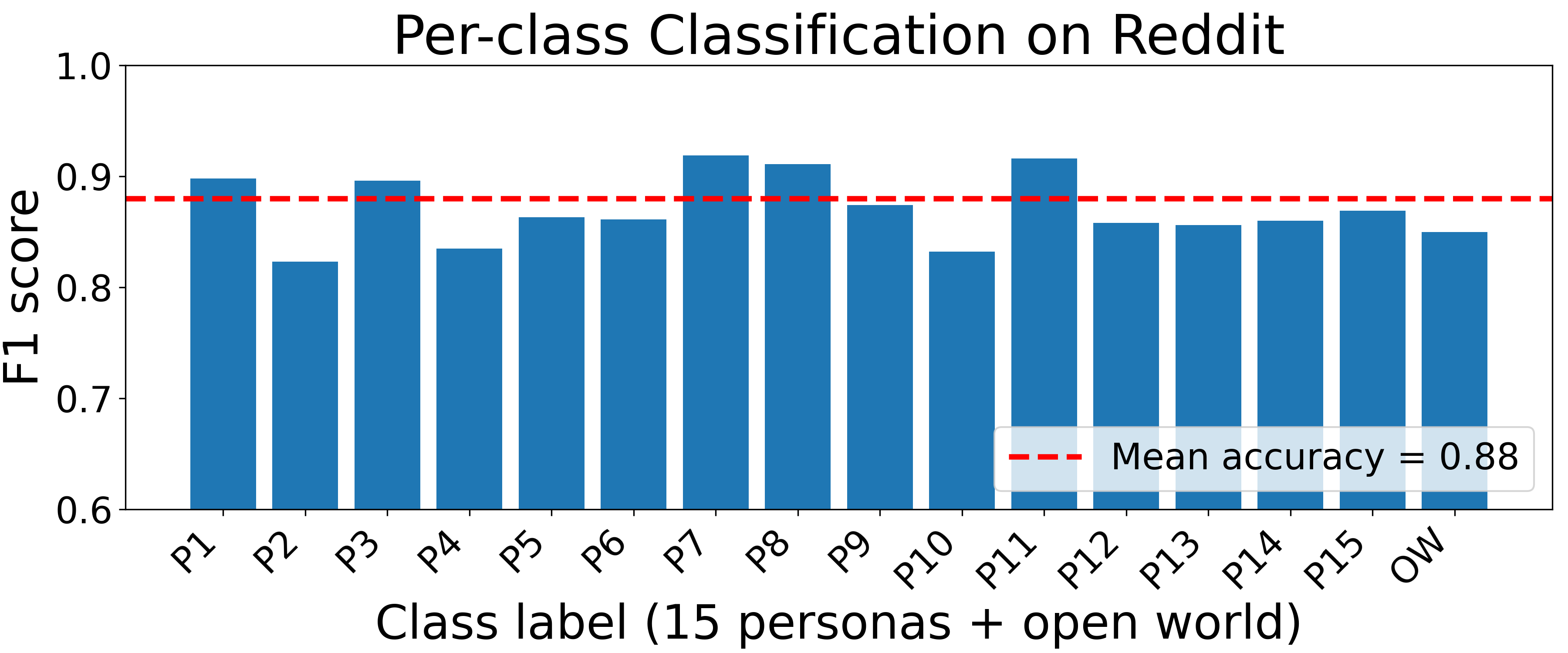}
    \caption{Reddit}
  \end{subfigure}

  \vspace{0.5em}

  \begin{subfigure}{0.19\textwidth}
    \centering
    \includegraphics[width=\linewidth]{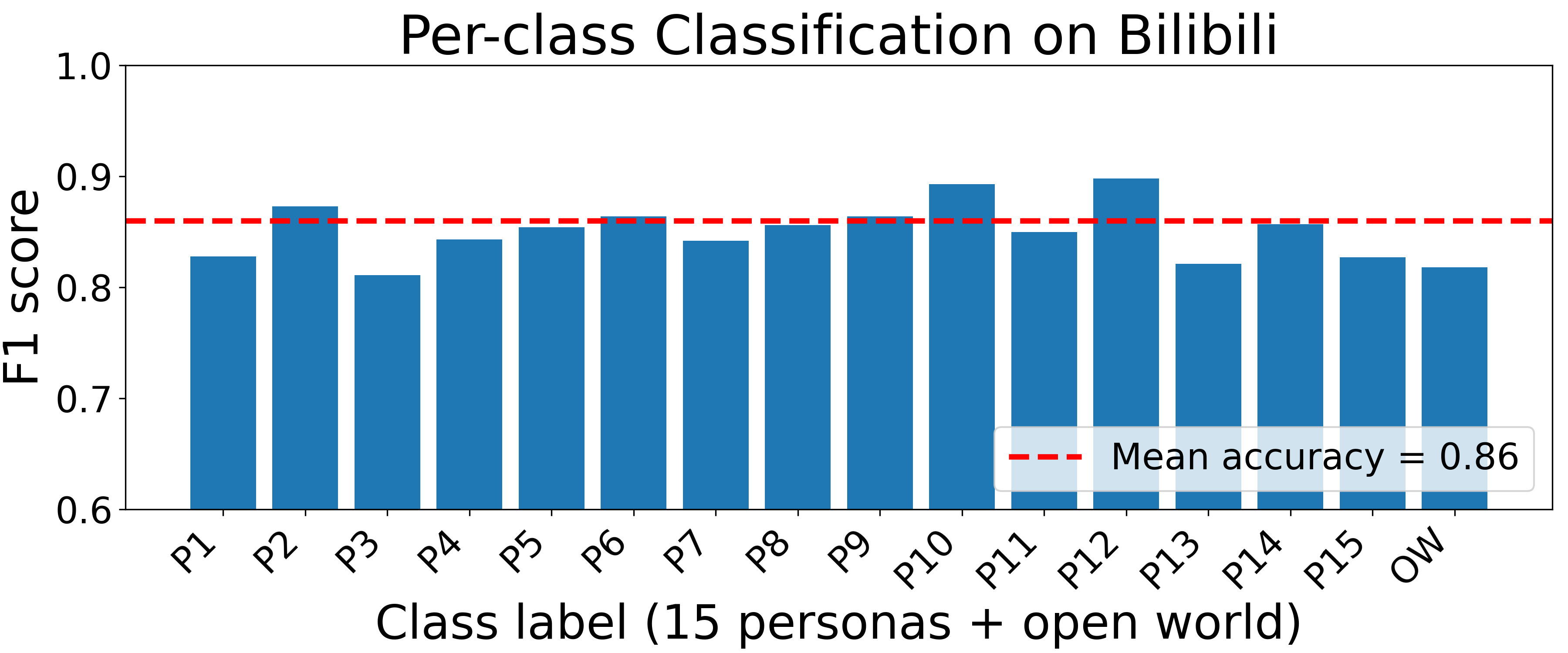}
    \caption{Bilibili}
  \end{subfigure}
  \hfill
  \begin{subfigure}{0.19\textwidth}
    \centering
    \includegraphics[width=\linewidth]{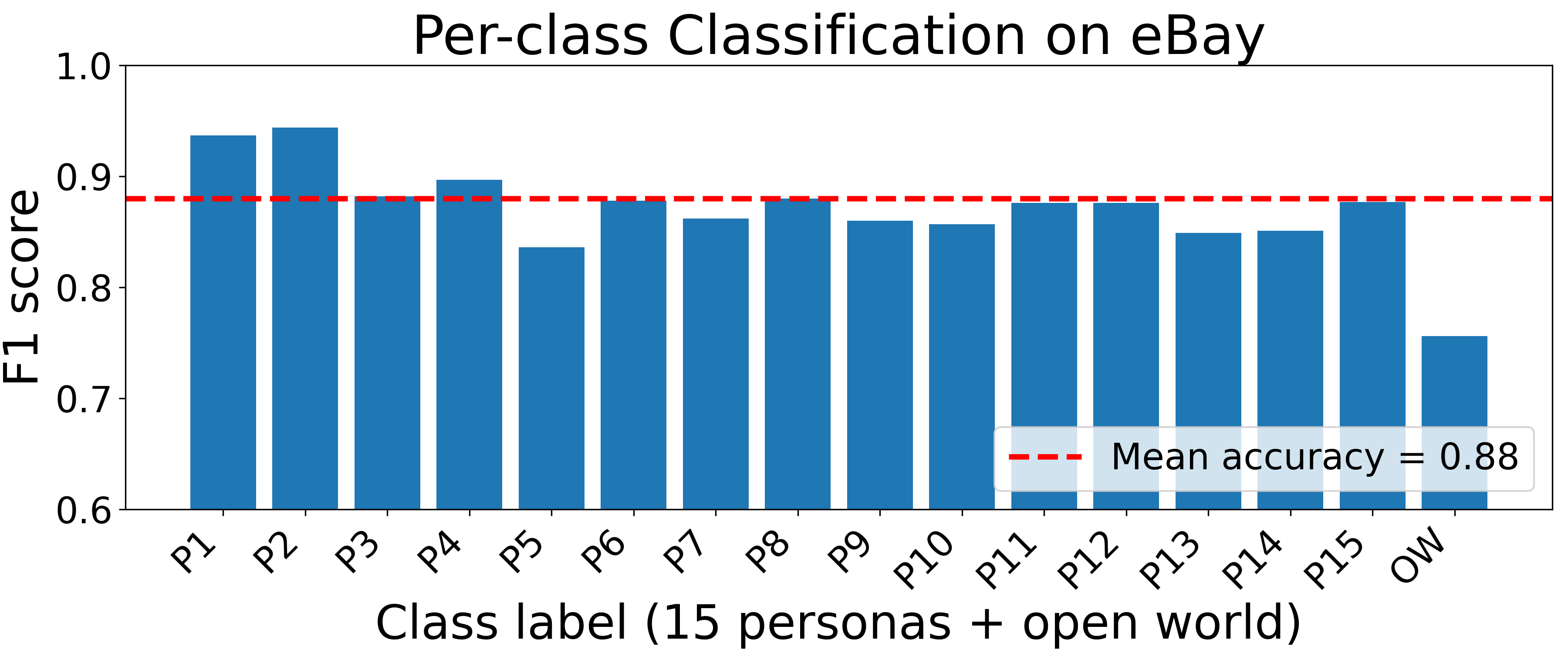}
    \caption{eBay}
  \end{subfigure}
  \hfill
  \begin{subfigure}{0.19\textwidth}
    \centering
    \includegraphics[width=\linewidth]{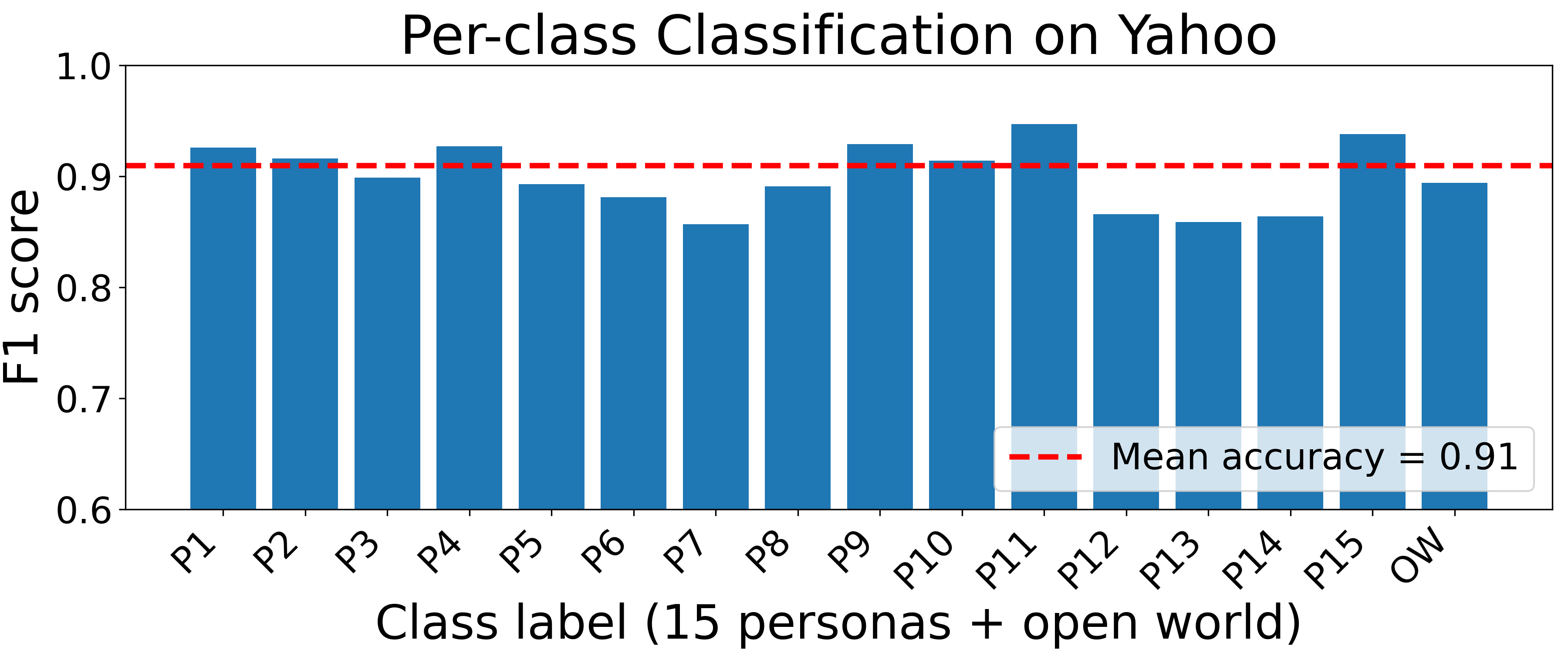}
    \caption{Yahoo}
  \end{subfigure}
  \hfill
  \begin{subfigure}{0.19\textwidth}
    \centering
    \includegraphics[width=\linewidth]{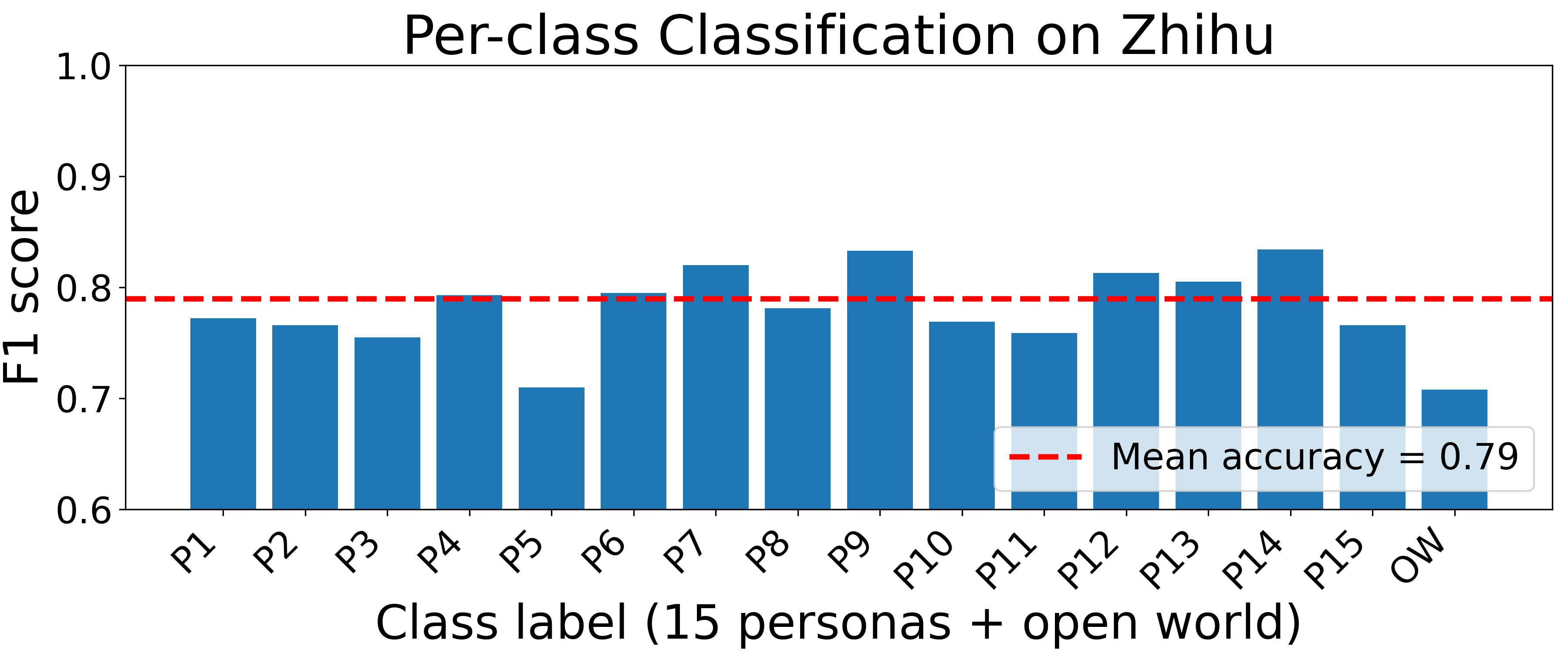}
    \caption{Zhihu}
  \end{subfigure}
  \hfill
  \begin{subfigure}{0.19\textwidth}
    \centering
    \includegraphics[width=\linewidth]{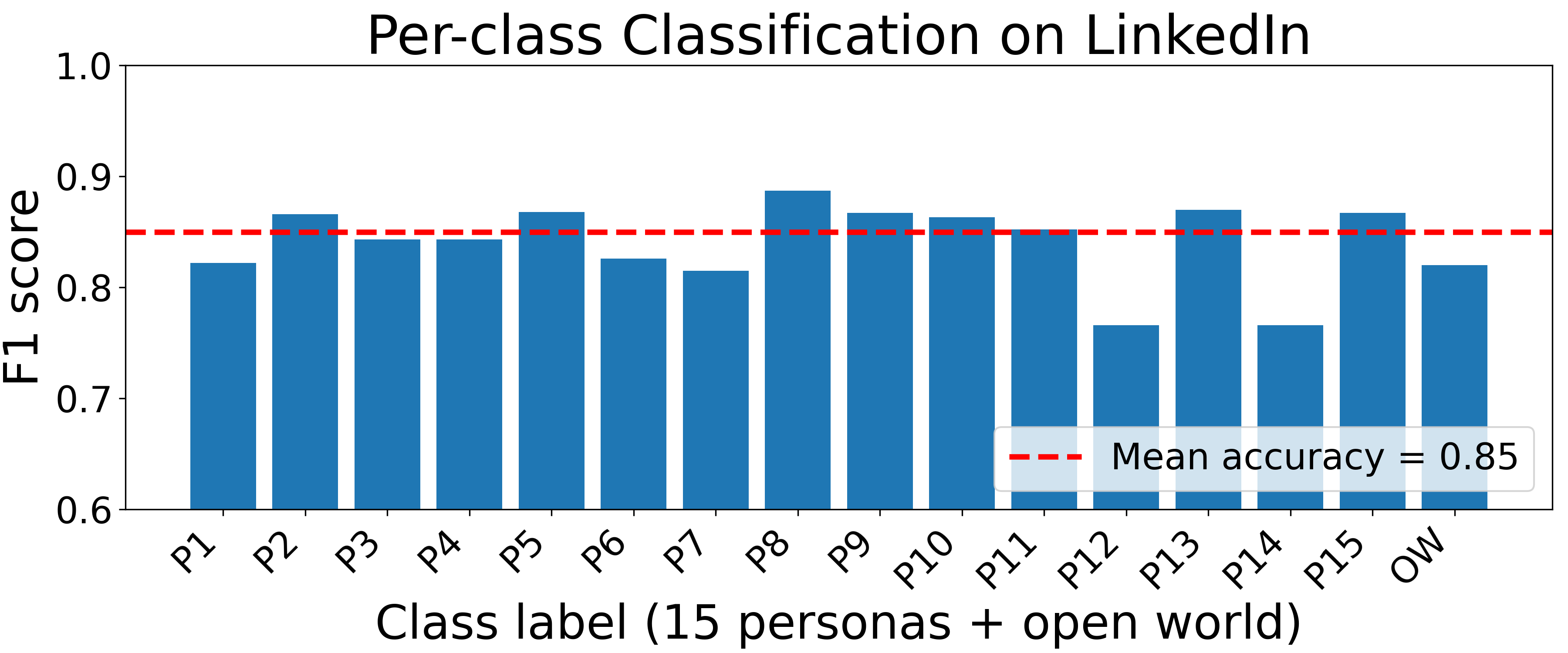}
    \caption{LinkedIn}
  \end{subfigure}

  \caption{Per-site persona fingerprinting: macro-\(F_1\) for each of the fifteen canonical personas and the \(\mathsf{OW}\) label on all ten websites. Each subfigure corresponds to one site, with the red horizontal line indicating overall persona accuracy on that site. Despite observing only 1,000-packet windows and no application payload, the attacker can reliably distinguish personas within each website.}
  \label{fig:persona-per-site}
\end{figure*}

Figure~\ref{fig:persona-per-site} summarizes per-site persona performance. For the first five sites, overall persona accuracy is high: around \(92\%\) on Amazon, \(75\%\) on CNN, \(88\%\) on Yelp, \(86\%\) on YouTube, and \(88\%\) on Reddit. The bar plots show that most personas achieve macro-\(F_1\) well above \(0.8\), with only a few borderline cases scoring lower when their browsing styles are intentionally designed to be similar.

The second row extends the analysis to five additional sites: Bilibili, eBay, Yahoo, Zhihu, and LinkedIn. Even on these heterogeneous platforms---ranging from video streaming (Bilibili) and Q\&A (Zhihu) to global e-commerce (eBay), portals (Yahoo), and professional networking (LinkedIn)---persona classification remains strong: average accuracy is around \(86\%\) for Bilibili, \(88\%\) for eBay, \(91\%\) for Yahoo, \(79\%\) for Zhihu, and \(85\%\) for LinkedIn. Personas that differ in exploration depth and topical focus (e.g., ``deep researcher'' vs.\ ``casual browser'') are consistently easy to distinguish, whereas intentionally similar personas are harder.

\subsection{Open-World Misattribution Risk}
\label{sec:evaluation-ow}

Per-site results in Figure~\ref{fig:persona-per-site} include the \(\mathsf{OW}\) class, which reflects whether the model can recognize browsing styles outside the canonical persona set. However, for persona inference the most consequential failure mode is often \emph{misattribution}: when an unseen persona is mapped to a specific canonical persona. This differs in an important way from open-world WFP. In open-world website fingerprinting, misclassifying an unmonitored site as a monitored one is typically interpreted as a drop in attack quality (and is often summarized via TPR/FPR). In contrast, \emph{persona} open-world inference can remain \emph{actionable} even when it is not perfectly correct: unseen personas are not arbitrary, and many naturally resemble one or more canonical personas along behavioral dimensions (e.g., exploration depth, scrolling cadence, or reliance on search vs.\ recommendations). Thus, collapsing an unseen persona into a nearby canonical persona can still provide a useful, coarse-grained behavioral profile for downstream profiling, even if the exact persona label is wrong. For this reason, analyzing open-world \emph{misattribution risk} is essential for understanding the privacy impact of persona fingerprinting.

To quantify this behavior, we go beyond macro-\(F_1\) and measure two attacker-relevant quantities for \(\mathsf{OW}\). \textbf{MisAttr@OW} is the fraction of \(\mathsf{OW}\) windows predicted as any \emph{canonical} persona (i.e., \(100-\mathrm{Recall}(\mathsf{OW})\)), capturing how often unseen behaviors are forced into a concrete persona label rather than rejected as ``other.'' \textbf{Top-3 Share} is the fraction of misattributed \(\mathsf{OW}\) windows concentrated in the top-3 predicted canonical personas; higher values indicate that misattributions collapse into a small number of stable persona profiles, which can be particularly damaging for profiling.

\begin{table}[t]
\centering
\small
\setlength{\tabcolsep}{5.5pt}
\resizebox{\linewidth}{!}{
\begin{tabular}{lccccc}
\toprule
\textbf{Site} & \textbf{OW Prec.} & \textbf{OW Rec.} & \textbf{OW F1} & \textbf{MisAttr@OW} & \textbf{Top-3 Share} \\
\midrule
Amazon   & 70.8 & 73.4 & 72.1 & 26.6 & 54.2 \\
YouTube  & 66.1 & 69.7 & 67.9 & 30.3 & 58.7 \\
Reddit   & 64.9 & 61.8 & 63.3 & 38.2 & 62.5 \\
CNN      & 62.3 & 60.4 & 61.3 & 39.6 & 60.1 \\
Yelp     & 71.6 & 66.2 & 68.8 & 33.8 & 49.4 \\
Bilibili & 59.7 & 57.1 & 58.4 & 42.9 & 66.8 \\
eBay     & 68.9 & 72.0 & 70.4 & 28.0 & 52.7 \\
Yahoo    & 63.8 & 67.6 & 65.6 & 32.4 & 55.9 \\
Zhihu    & 61.5 & 56.9 & 59.1 & 43.1 & 69.3 \\
LinkedIn & 67.2 & 70.5 & 68.8 & 29.5 & 57.0 \\
\midrule
\textbf{Avg.} & \textbf{65.7} & \textbf{65.6} & \textbf{65.6} & \textbf{34.4} & \textbf{58.7} \\
\bottomrule
\end{tabular}
}
\vspace{2pt}
\caption{\textbf{Open-world misattribution risk (per-site).} \textbf{MisAttr@OW} is the fraction of OW windows predicted as any \emph{canonical} persona (i.e., $100-\mathrm{Recall}(\mathsf{OW})$). \textbf{Top-3 Share} is the fraction of misattributed OW windows concentrated in the top-3 predicted canonical personas (higher implies errors collapse to a few persona labels). Values are percentages.}
\label{tab:ow-misattribution}
\end{table}

Table~\ref{tab:ow-misattribution} reports \(\mathsf{OW}\) precision/recall/\(F_1\) and our misattribution metrics. On average, \(\mathsf{OW}\) achieves \(65.6\%\) \(F_1\), indicating that the classifier often recognizes when traffic does not match any canonical persona. Yet the remaining errors are substantial: \textbf{MisAttr@OW averages \(34.4\%\)}, meaning roughly one-third of unseen-persona windows are assigned to a specific canonical persona rather than labeled as \(\mathsf{OW}\). Misattribution varies by platform, ranging from \(\approx 26\%\!-\!28\%\) on Amazon/eBay to \(\approx 43\%\) on Bilibili/Zhihu, suggesting that some sites induce persona-style traffic patterns that are easier to confuse with canonical profiles under packet-metadata observations.

Moreover, misattributions are often \emph{concentrated}: \textbf{Top-3 Share averages \(58.7\%\)} and exceeds \(65\%\) on Bilibili and Zhihu. This ``profiling collapse'' means that when \(\mathsf{OW}\) is misclassified, the resulting labels frequently fall into a small set of canonical personas---precisely the outcome that can still be useful for coarse-grained behavioral profiling. These findings motivate treating open-world persona inference as more than a standard classification problem: the structure of misattribution is itself an important privacy signal, complementary to overall macro-\(F_1\).

\subsection{Cross-Site Persona Generalization}
\label{sec:evaluation-cross-site}

The previous subsection studied persona fingerprinting within a single website at a time. We now ask whether persona signatures persist when traffic from multiple sites is mixed together, and whether a single persona classifier can generalize across heterogeneous sites.

\begin{figure}[t]
  \centering
  \includegraphics[width=\linewidth]{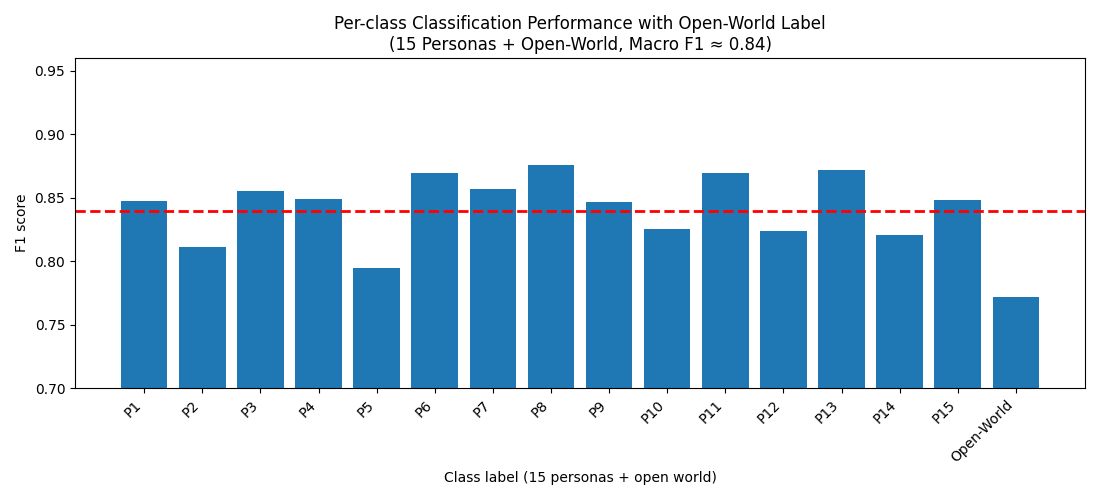}
  \caption{Global persona fingerprinting in an open-world setting: macro-\(F_1\) for the fifteen canonical personas and the \(\mathsf{OW}\) label when combining traffic from multiple websites. Overall persona accuracy remains high (around \(84\%\)), indicating that the learned fingerprints transfer across sites under heterogeneous modern-web traffic.}
  \label{fig:persona-global-openworld}
\end{figure}

We train a single persona classifier on mixed traffic from multiple sites, again using only 1{,}000-packet windows as input. Figure~\ref{fig:persona-global-openworld} shows the resulting macro-\(F_1\) for all fifteen personas plus \(\mathsf{OW}\). Despite additional heterogeneity induced by mixing sites, overall persona accuracy remains around \(84\%\), and most individual personas maintain macro-\(F_1\) scores comparable to their per-site counterparts.

We also compare a closed-set variant (predicting only the 15 canonical personas) with the full open-world classifier (15 personas plus \(\mathsf{OW}\)). Aggregating across sites, the closed-set model achieves roughly \(82\%\) accuracy, while the open-world model reaches about \(80\%\). Thus, introducing an explicit \(\mathsf{OW}\) label---which forces the model to absorb diverse unseen-persona behaviors into a single ``other'' class---reduces accuracy only slightly. Mixed-site success suggests that persona cues are not narrowly tied to a specific site's layout, but reflect interaction-driven traffic rhythms that persist across modern websites.

\subsection{Attack Scaling with More Persona-Labeled Traffic}
\label{sec:evaluation-scaling}

A central motivation for using LLM agents is that they make persona-labeled traffic \emph{scalable} (Section~\ref{sec:methodology}), which can directly strengthen attackers who train persona inference models. To quantify this effect, we vary the training budget by subsampling the number of labeled windows per persona and evaluate mixed-site open-world persona inference.

\begin{table}[t]
\centering
\small
\setlength{\tabcolsep}{6.5pt}
\resizebox{\linewidth}{!}{
\begin{tabular}{rcc}
\toprule
\textbf{Train windows / persona} & \textbf{10-site Mixed (OW) Acc. (\%)} & \textbf{5-site Mixed (OW) Acc. (\%)} \\
\midrule
500  & 55.0$\pm$1.0 \; (53.8--55.6) & 60.5$\pm$0.8 \; (59.6--61.1) \\
1000 & 65.0$\pm$1.0 \; (63.9--65.7) & 69.8$\pm$0.8 \; (68.9--70.4) \\
2000 & 76.0$\pm$0.7 \; (75.2--76.4) & 79.0$\pm$0.7 \; (78.2--79.4) \\
5000 & 84.0$\pm$0.8 \; (83.1--84.5) & 86.2$\pm$0.8 \; (85.3--86.7) \\
\bottomrule
\end{tabular}
}
\vspace{2pt}
\caption{\textbf{Attack scaling with more persona-labeled traffic (open-world).} Mixed-site persona inference accuracy improves rapidly at small budgets and continues to rise from 2,000 to 5,000 windows/persona. Results are mean$\pm$std over 3 seeds with min--max in parentheses.}
\label{tab:scaling-ow-acc}
\end{table}

Table~\ref{tab:scaling-ow-acc} shows a steep scaling curve. With only 500 training windows per persona, 10-site mixed open-world accuracy is \(55.0\%\), but it increases to \(65.0\%\) at 1{,}000 windows/persona and \(76.0\%\) at 2{,}000. Even beyond that, accuracy continues to rise to \(84.0\%\) at 5{,}000 windows/persona, indicating that additional labeled traffic remains beneficial at larger budgets. The gains are consistent across random seeds (small standard deviations and narrow min--max ranges), suggesting that improvements are not driven by a particular split or initialization.

We also report a 5-site mixed setting, which is consistently easier than 10 sites (e.g., \(60.5\%\) vs.\ \(55.0\%\) at 500 windows/persona), but the gap narrows as the budget grows (e.g., \(86.2\%\) vs.\ \(84.0\%\) at 5{,}000 windows/persona). This pattern supports the attacker-centric interpretation: heterogeneous sites increase the complexity of mixed-site inference at small budgets, yet sufficient persona-labeled data can overcome much of this heterogeneity. Overall, these results directly substantiate the ``scalability is pivotal'' thesis: once persona-labeled traffic becomes easy to produce at scale, metadata-based persona inference can improve rapidly and reach high accuracy.

\subsection{Representation Probing and Joint Multi-Task Training}
\label{sec:evaluation-joint}

We next investigate how much persona information is already encoded in a standard WFP encoder, and whether an attacker can easily upgrade a site-only model into a strong persona fingerprinting model by joint training.

\subsubsection{Probing site-only encoders for persona leakage}

We perform a representation probing experiment in two stages. First, we train the site-only WFP baseline from Section~\ref{sec:evaluation-site-only} until convergence. Second, we \emph{freeze} the encoder \(h_{\theta}\) and train a \emph{lightweight MLP probe} \(f_{\mathsf{probe}}\) (Section~\ref{sec:probing}) on top of it to predict persona labels. For comparison, we train an identical MLP probe on top of a randomly initialized frozen encoder \(h_{\text{rand}}\) that has never seen any packet data.

\begin{figure}[t]
  \centering
  \begin{subfigure}{0.49\linewidth}
    \centering
    \includegraphics[width=\linewidth]{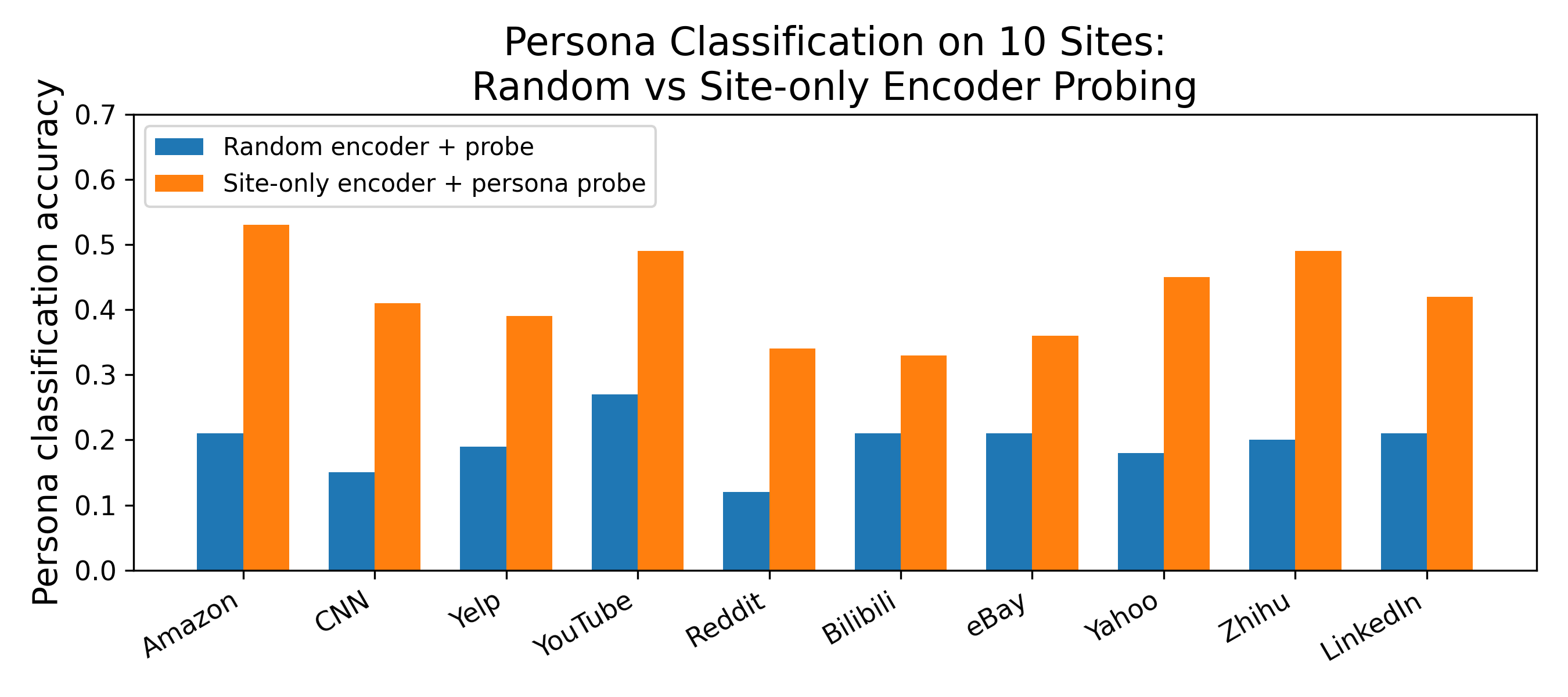}
    \caption{Accuracy per site}
  \end{subfigure}
  \hfill
  \begin{subfigure}{0.48\linewidth}
    \centering
    \includegraphics[width=\linewidth]{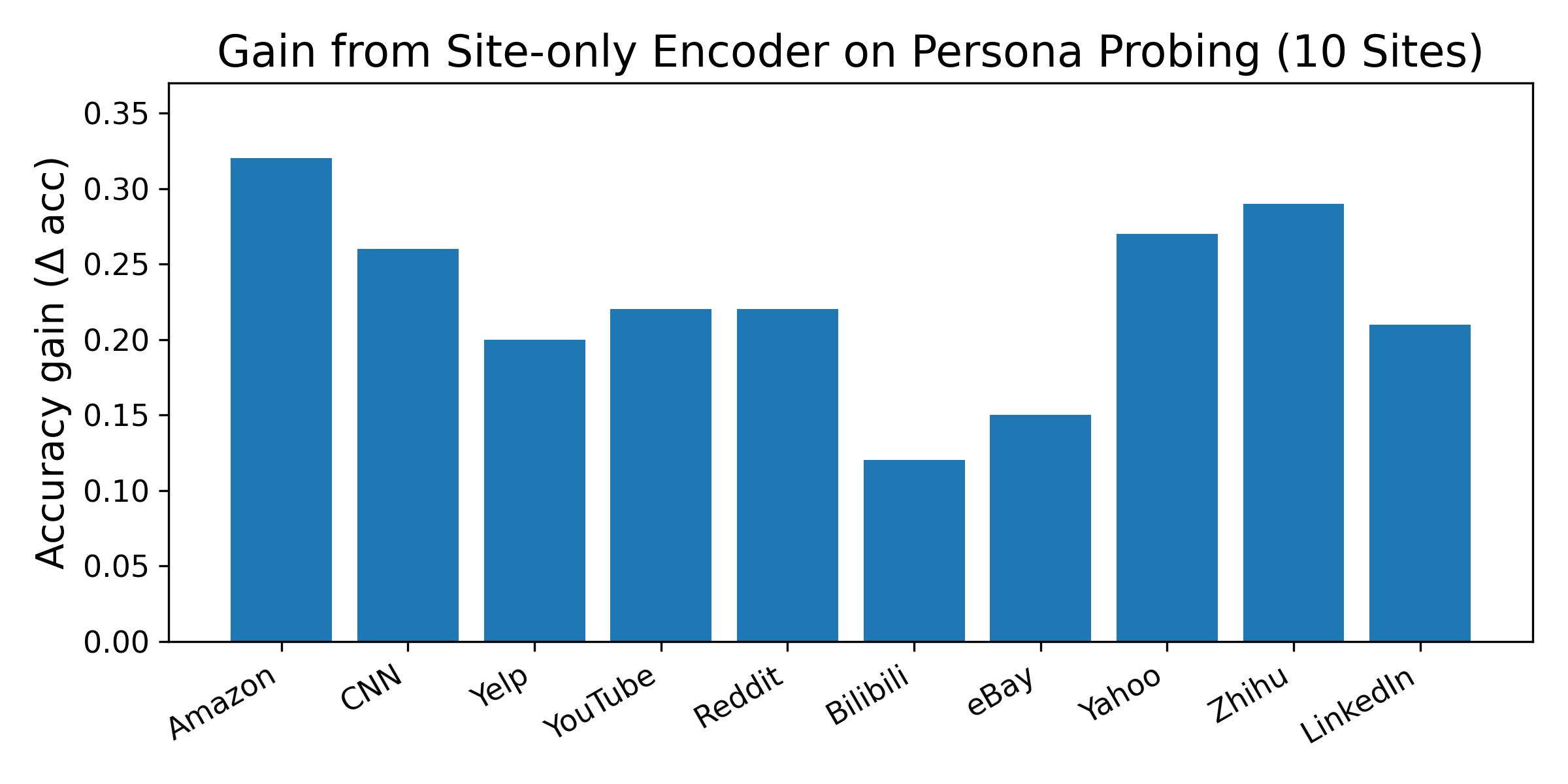}
    \caption{Accuracy gain}
  \end{subfigure}
  \caption{Representation probing: persona classification accuracy using a lightweight MLP probe on (i) a random encoder and (ii) the site-only encoder trained for WFP. Across all sites, the site-only encoder provides a substantial gain (typically \(+20\%\!-\!30\%\) absolute), indicating that standard WFP training already implicitly encodes readily decodable persona information.}
  \label{fig:persona-probing}
\end{figure}

Figure~\ref{fig:persona-probing}(a) shows persona accuracy for both probes on each site. On the original five sites, accuracy on the random encoder is very low (e.g., around \(21\%\) on Amazon, \(15\%\) on CNN, \(19\%\) on Yelp, \(27\%\) on YouTube, and \(12\%\) on Reddit), as expected for a high-cardinality label space without meaningful representations. In contrast, the probe on the site-only encoder attains much higher accuracy---roughly \(53\%\), \(41\%\), \(39\%\), \(49\%\), and \(34\%\) on these sites, respectively---despite the encoder never being trained directly on persona labels. Similar gains appear on the additional five sites, as summarized in Figure~\ref{fig:persona-probing}(b), where the per-site accuracy improvement is typically between \(20\%\) and \(30\%\) absolute. These results show that an attacker can often extract persona signals from a WFP encoder with minimal additional modeling, making persona leakage plausible even when the original intent was only site identification.

\subsubsection{Joint modeling of websites and personas}

We now explicitly train a joint model that predicts both websites and personas. We reuse the shared encoder \(h_{\theta}\), attach two heads \(f_{\mathsf{site}}\) and \(f_{\mathsf{pers}}\), and optimize the combined loss
\[
  L_{\text{joint}} \;=\; L_{\mathsf{site}} \;+\; \lambda \, L_{\mathsf{pers}},
\]
where \(\lambda \geq 0\) controls the strength of persona supervision. When \(\lambda = 0\), the model reduces to the site-only baseline; as \(\lambda\) increases, the encoder is progressively steered toward persona-discriminative features.

\begin{figure}[t]
  \centering
  \includegraphics[width=\linewidth]{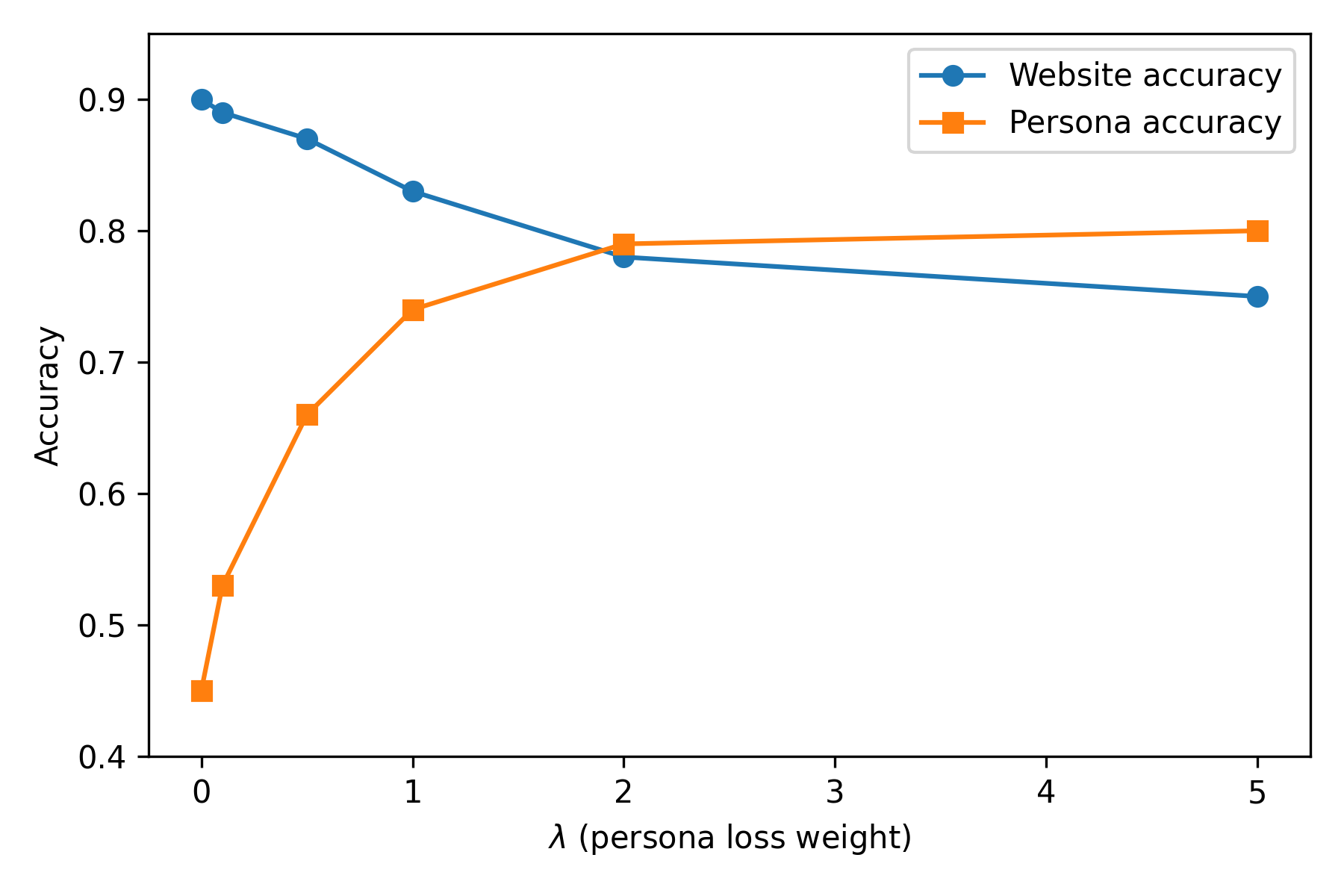}
  \caption{Joint multi-task trade-off between website and persona accuracy as a function of the persona loss weight \(\lambda\). Increasing \(\lambda\) dramatically improves persona accuracy (from \(\approx 45\%\) to \(\approx 80\%\)) while only gradually reducing website accuracy (from \(\approx 90\%\) to \(\approx 75\%\)).}
  \label{fig:joint-lambda-sweep}
\end{figure}

Figure~\ref{fig:joint-lambda-sweep} illustrates the trade-off on the mixed-site persona task. At \(\lambda = 0\), the model behaves like the pure WFP baseline: website accuracy is about \(90\%\), while persona accuracy lags behind at roughly \(45\%\). As we increase \(\lambda\) to \(0.5\) and \(1.0\), persona accuracy rises sharply (to around \(66\%\) and \(74\%\), respectively), while website accuracy remains above \(80\%\). For larger values such as \(\lambda = 2.0\) and \(5.0\), persona accuracy saturates near \(80\%\), whereas website accuracy declines to around \(75\%\). In other words, there exists a broad regime where joint training yields large gains in persona fingerprinting at only modest cost to WFP performance. Combined with the scaling results in Section~\ref{sec:evaluation-scaling}, this suggests a practical attacker trajectory: once persona-labeled data becomes available at scale, upgrading a WFP pipeline to include persona inference can be straightforward and effective.

\subsection{Persona Consistency and Behavioral Diversity}
\label{sec:evaluation-consistency}

Finally, we evaluate whether the LLM-generated browsing traces behave like coherent personas rather than degenerate scripts. Recall that Section~\ref{sec:quality} introduced behavioral diagnostics including the persona consistency score \(C(p)\) and site coverage; here we report both and compare two data-generation configurations: (i) a baseline without persona-aware prompting or exploration mechanisms, and (ii) our full multi-agent pipeline with rich persona prompts, history-aware commands, and explicit encouragement to explore multiple page types.

\begin{figure}[t]
  \centering
  \includegraphics[width=\linewidth]{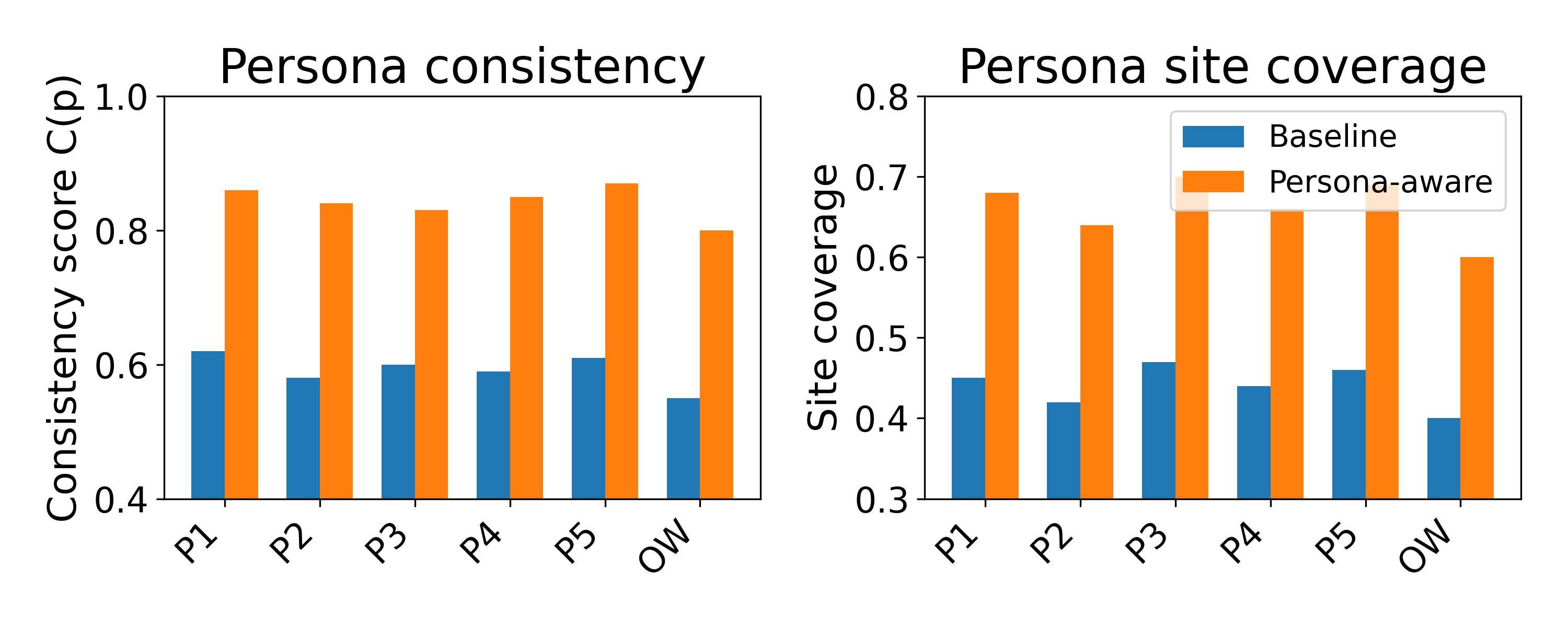}
  \caption{Behavioral diagnostics for a subset of personas (P1--P5 and \(\mathsf{OW}\)), comparing a baseline generation strategy with our persona-aware multi-agent pipeline. Persona-aware generation substantially increases both consistency \(C(p)\) and site coverage, indicating that LLM agents behave more like coherent, diverse personas rather than degenerate scripts.}
  \label{fig:persona-consistency-coverage}
\end{figure}

As shown in Figure~\ref{fig:persona-consistency-coverage}, persona-aware generation increases consistency scores from roughly \(0.58\!-\!0.62\) in the baseline to around \(0.83\!-\!0.87\), meaning that an external judge now finds behaviors to align much more closely with intended persona descriptions. At the same time, site coverage rises from about \(0.40\!-\!0.47\) to \(0.64\!-\!0.70\), suggesting that personas explore a broader set of pages and interaction patterns instead of repeating a narrow script. Crucially, these improvements in consistency and coverage co-exist with the high persona classification accuracy observed in Sections~\ref{sec:evaluation-per-site-persona} and~\ref{sec:evaluation-cross-site}, supporting the claim that learned persona fingerprints reflect robust behavioral patterns rather than brittle artifacts of under-specified data generation.

\section{Related Work}
\label{sec:related}

\textbf{Website fingerprinting models.}
Early work on website fingerprinting (WFP) uses hand-crafted features and
classical machine learning in small closed worlds.  Subsequent deep
models based on CNN and transformer architectures\cite{10.1145/3576915.3623107,Bhat_2019,10.1145/3485832.3485891,10.1145/3696410.3714578,sirinam2018deepfingerprintingunderminingwebsite} operate directly on
packet sequences and achieve near-perfect closed-world accuracy under
synthetic conditions.  More recent approaches explore robustness via
contrastive learning\cite{chen2020simpleframeworkcontrastivelearning}, domain transfer\cite{song2024seamlesswebsitefingerprintingmultiple}, and adversarial augmentation\cite{zhang2019adversarialautoaugment}, but
still largely assume short, scripted sessions and focus on site-level
classification\cite{10.1145/2660267.2660362,203876,247654,10.1145/3658644.3690211}.  Our work follows this line of deep WFP models, but
evaluates them under long, multi-step, high-entropy browsing with
persona labels and an open-world label space.

\textbf{Behavioral and persona fingerprinting.}
Beyond WFP, a broad literature studies\cite{Olejnik2012WhyJC,255662} how users and devices can be
identified from behavioral traces, including network traffic patterns,
interaction timing, and other implicit signals.  These studies show that
even when content is hidden or identifiers are removed, stable
individual or group-specific patterns often remain.  In contrast, most
existing work assumes access to rich application- or user-level logs.
We instead study persona fingerprints at the network layer, showing that
coarse, encrypted packet sequences already carry enough structure to
distinguish behavioral personas in a multi-site, open-world setting.

\textbf{LLM-based agents for web interaction.}
Recent systems use large language models as agents that plan, remember,
and invoke tools over long interaction horizons.  Browser-interacting
agents\cite{pan2024webcanvasbenchmarkingwebagents,drouin2024workarenacapablewebagents,dechezelles2025browsergymecosystemwebagent,deng2023mind2webgeneralistagentweb} extend this idea to real user interfaces, combining
vision-language reasoning with keyboard and mouse control to complete
web tasks.  Prior work in this space typically optimizes task success or
instruction following, without examining the resulting network traffic.
Our approach repurposes such agents as a controllable traffic generator:
a persona-conditioned decision agent drives a computer-use agent in a
real browser, producing realistic, high-entropy sessions whose encrypted
packet traces we use to study website and persona fingerprinting.

\section{Future Work}
\label{sec:future-work}

Building on our empirical study of persona fingerprints in encrypted
web browsing, we highlight several directions for future work.

\paragraph{Larger-scale and longer-term traces.}
An immediate next step is to validate our findings on larger and
longer-term datasets that combine synthetic and real users.  This
includes studying how persona signatures behave across days or weeks,
how stable they remain under changing network conditions and devices,
and how to calibrate models trained primarily on LLM-generated traces
to real-world traffic via domain adaptation or transfer learning.
Extending our analysis from 1{,}000-packet windows within a single
connection to richer session- and user-level timelines would also shed
light on how temporal aggregation affects persona leakage.

\paragraph{Persona-aware risk metrics and defenses.}
Our results suggest that existing encrypted-browsing systems should be
evaluated not only for website fingerprinting risk but also for
persona fingerprinting risk.  A promising direction is to develop
formal metrics and benchmarks that jointly track website accuracy,
persona accuracy, and utility, and to use them to co-design
\emph{persona-aware} defenses.  Such defenses may combine traffic
shaping with changes to browser and application behavior to reduce
learnable long-term behavioral patterns, rather than focusing solely
on obfuscating site identity.

\section{Conclusion}
\label{sec:conclusion-final}

This paper takes a first step toward systematically understanding
\emph{persona fingerprints in encrypted web browsing}. We consider a
realistic attacker who passively observes only encrypted traffic
\emph{metadata} during browsing sessions to modern websites---including
naturally embedded third-party resources such as ads---and extracts
fixed-length packet windows (Section~\ref{sec:representation}), with no
access to URLs, HTTP headers, or page content. Within this constrained
view, we ask what an attacker can infer not only about \emph{which
website} a user visits, but also \emph{how the user browses} in terms of
stable behavioral personas.

Our first contribution is to define and empirically measure persona
fingerprints across ten modern websites and sixteen labels (fifteen
canonical personas plus an Open-World class \(\mathsf{OW}\)). Using a
shared encoder \(h_{\theta}(X)\) and simple classification heads, we
show that persona classification achieves high accuracy within
individual websites and remains strong in mixed-site open-world
scenarios from 1{,}000-packet windows.

Second, we introduce a controllable multi-agent LLM framework for
generating persona-conditioned browsing traffic. A decision agent,
driven by persona-specific system prompts and history summaries,
produces high-level commands that a computer-use agent executes in a
real browser. We propose behavioral diagnostics---including a persona
consistency score \(C(p)\) and coverage/diversity measures---and show
that a persona-aware pipeline improves both consistency and coverage
over naive generation, yielding a practical, reproducible source of
persona-labeled traffic for privacy studies.

Third, we show that existing website fingerprinting encoders already
exhibit substantial \emph{latent persona leakage}. A site-only encoder
trained without persona labels learns representations from which a
lightweight MLP probe recovers persona information far above chance, and
joint multi-task training with modest weight \(\lambda\) maintains strong
website accuracy while raising persona accuracy to around \(80\%\).
Crucially, this risk scales with data: mixed-site open-world persona
accuracy improves rapidly with more persona-labeled windows per persona,
underscoring that scalable traffic generation can materially strengthen
metadata-based persona inference.

\section{Ethical Considerations}
\label{sec:ethics}


Our study is a measurement and risk-characterization effort on privacy leakage from encrypted web traffic metadata. 
We do not recruit human participants, run user studies, collect surveys, or access any content-level information; instead, our datasets are generated by an LLM-driven, computer-use browsing agent interacting with real websites under strict safety constraints. In addition, any real-human traffic used in our experiments comes from public datasets and is used strictly in accordance with their stated terms and usage requirements. In particular, we capture only packet-level metadata and do not collect URLs, HTTP headers, cookies, credentials, or any personally identifying information. To minimize potential harm, our automated browsing avoids login/payment flows and does not submit forms with sensitive information. Consistent with responsible disclosure norms, we avoid releasing any end-to-end attack tool; any shared artifacts (e.g., persona definitions, evaluation scripts, and aggregate results) are designed to support reproducibility without enabling abuse.

\bibliographystyle{plain}
\bibliography{usenix2024_SOUPS}

\end{document}